\documentclass[amsmath,amssymb,showpacs,superscriptaddress]{revtex4}

\usepackage{bm,amsmath,amsthm,amssymb,hyperref,graphicx}

\newtheorem{thm}{Theorem}[section]
\newtheorem{lem}{Lemma}[section]

\newcommand{\be}{\begin{equation}}
\newcommand{\ee}{\end{equation}}
\newcommand{\bea}{\begin{eqnarray}}
\newcommand{\eea}{\end{eqnarray}}
\newcommand{\6}{\partial}
\newcommand{\fa}{\Psi}
\newcommand{\fad}{\Psi^\dagger}
\newcommand{\lam}{\lambda}
\newcommand{\ra}{\rangle}
\newcommand{\la}{\langle}
\newcommand{\inti}{\int_{-\infty}^{+\infty}}

\newcommand{\thet}{\phi}

\begin{document}

\title{One-Dimensional Impenetrable Anyons in Thermal Equilibrium. IV.
Large Time and Distance Asymptotic Behavior of the Correlation Functions}

\author{Ovidiu I. P\^{a}\c{t}u}

\affiliation{Fachbereich C – Physik, Bergische Universit\"at  Wuppertal,
 Wuppertal 42097, Germany }
\affiliation{Institute for Space Sciences,
 Bucharest-M\u{a}gurele, R 077125, Romania}

\author{Vladimir E. Korepin}

\affiliation {C.N. Yang Institute  for Theoretical Physics, State
University of New York at Stony Brook, Stony Brook, NY 11794-3840,
USA }

\author{Dmitri V. Averin}
\affiliation{Department of Physics and Astronomy, State University
of New York at Stony Brook, Stony Brook, NY 11794-3800, USA }

\begin{abstract}
This work presents the derivation of the large time and distance asymptotic behavior of the field-field correlation functions of impenetrable one-dimensional anyons at finite temperature. In the appropriate limits of the statistics parameter, we recover the well-known results for
impenetrable bosons and free fermions. In the low-temperature (usually expected to be the ``conformal'') limit, and for all values of the statistics parameter away from the bosonic point, the leading term in the correlator does not agree with the prediction of the conformal field theory, and is determined by the singularity of the density of the single-particle states at the bottom of the single-particle energy spectrum.
\end{abstract}

\maketitle

\section{Introduction}

This is the last article in the series of papers
\cite{PKA2,PKA3,PKA5} in which we study rigorously the large time
and distance asymptotic  behavior of the temperature dependent
field-field correlation functions of one-dimensional impenetrable
anyons. In this work, we present the derivation of the final results
for the asymptotics of the time-dependent correlation functions. As
in the case of  ``static'' (same-time) correlators, for which the
asymptotic behavior was computed in \cite{PKA5}, the starting point
of our  analysis is the determinant representation for the
correlators found in \cite{PKA2,PKA3}. With the help of this
representation, we are able to derive a system of differential
equations for the correlators, which is the same as the one for
impenetrable bosons \cite{IIKS,KBI}, but with different initial
conditions. The asymptotic behavior of the correlators is computed
then by solving the matrix Riemann-Hilbert problem that is
associated with the obtained system of the differential equations.
The most striking feature of the time-dependent asymptotics found in
this work is the fact that its leading term is non-conformal. It
contradicts the predictions of the conformal field theory (or,
equivalently, bosonization) that were derived for the
one-dimensional anyons in \cite{CM,PKA}. This is in contrast to the
static correlators which do agree with the conformal field theory.

The model of impenetrable anyons considered in our series of papers
is arguably the simplest physical model of one-dimensional particles
with fractional exchange statistics, and is the anyonic
generalization of the impenetrable Bose gas first studied by
Girardeau \cite{Gir1}. Despite its simplicity, the model is closely
related to the realistic models of transport of anyonic
quasiparticles of the fractional quantum Hall effect \cite{AN}. The
model of impenetrable anyons can also be viewed as the
infinite-repulsion limit of a more general model of anyons with
$\delta$-function interaction of finite strength, called the
Lieb-Liniger gas of anyons, and suggested in \cite{Kundu}.
Introduction of the fractional exchange statistics in one dimension
requires additional convention for the direction of the
particle-particle exchanges \cite{AN,PKA}. This implies that for
finite anyon-anyon interaction, the anyonic wavefunction is
discontinuous at the coincident particle coordinates, the fact that
makes the physical interpretation of the finite-interaction case
difficult. Nevertheless, the Lieb-Liniger gas of anyons can be
well-defined mathematically, and has received considerable attention
in the last few years. As a result of these efforts, we know the
Bethe Ansatz solution \cite{Kundu} of this model, the low-energy
properties and the connection with Haldane's \cite{Hald} fractional
exclusion statistics \cite{BGO,BG}, the thermodynamics \cite{BGH},
the ground-state properties \cite{HZC}, and the low-lying
excitations \cite{PKA}. Various techniques were used to study the
correlation functions (mostly for the physically-motivated
impenetrable case) such as: the Fisher-Hartwig conjecture
\cite{SSC,SC}, bosonization \cite{CM}, conformal field theory \cite{PKA},
numerical calculations \cite{C,GHC}, and the replica method
\cite{SC2}. The present paper, together with our previous papers
\cite{PKA2,PKA3,PKA5}, is devoted to the exact calculation of the
asymptotic behavior of the correlation functions using the
techniques developed for impenetrable bosons
\cite{IIK1,IIK2,IIK3,IIKS,IIKV,KBI}. It should be mentioned that
other models of the one-dimensional fractional exchange statistics
\cite{AOE1,OAE1,IT,LMP,BGK,HZC1,BCM} can also be found in the
literature. In particular, the quantum inverse scattering method
with anyonic grading was developed recently in \cite{BFGLZ}.

The main result obtained in this work is the large time and distance
asymptotics of the field-field correlator
$\la\fa(x_2,t_2)\fad(x_1,t_1) \ra_T$ of impenetrable anyons at
finite temperatures. This result can be expressed conveniently in
the rescaled variables [see Eq.~(\ref{rescaled})] in which
\be \la\fa(x_2,t_2)\fad(x_1,t_1)\ra=\sqrt{T} g(x,t,\beta,\kappa)\, ,
\label{corr}  \ee
and the function $g(x,t,\beta,\kappa)$ is defined below. To do this,
we need to introduce several quantities:
\be \label{int2}
C(x,t,\beta,\kappa)=\frac{1}{\pi}\inti|x-2t\lam|\ln|\varphi(\lam^2, \beta, \kappa)|\ d\lam\, ,\ \ \ \ I(\beta,\kappa)=\Im\left(\inti  \ln \varphi(\lam^2,\beta,\kappa)\ d\lam \right)\, ,
\ee
and
\be \label{int3}
\varphi(\lam^2,\beta,\kappa)=\frac{e^{\lam^2-\beta}-e^{i\pi\kappa}}{e^{ \lam^2 -\beta}+1}\, , \ee
where the branch of the logarithm is chosen so that
$\lim_{\lam\rightarrow \infty} \ln \varphi(\lam^2,\beta,\kappa)=0$.
With these definitions, our result for the function
$g(x,t,\beta,\kappa)$ can be stated as follows. In the large time
and distance limit: $x,t\rightarrow \infty$, with $x/t=const$, the
asymptotic behavior of the field-field correlation function
(\ref{corr}) is given by
\be \label{mainres} g(x,t,\beta,\kappa)=
t^{\nu^2/2}e^{C(x,t,\beta,\kappa)+ix I(\beta,\kappa)}[c_0 t^{-1/2-i\nu}e^{2it(\lam_s^2+\beta)}+
c_1 e^{2(\pm t\pi\kappa-ix\lam_0^{\mp})} +o(1/\sqrt{t} )] \, ,
\ee
where $\lam_s=-x/2t$ and $\nu=-\frac{1}{\pi}\ln| \varphi(\lam_s^2,
\beta,\kappa)|$. Other notations are:
\be \label{int5}
\lam_0^{\mp}=-\left(\beta+ \sqrt{\beta^2+\pi^2\kappa^2}\right)^{1/2} /\sqrt{2} \mp i\left(-\beta+ \sqrt{\beta^2 +\pi^2\kappa^2} \right)^{1/2} /\sqrt{2}\, ,
\ee
with $\kappa\in[0,1]$ being the statistics parameter: $\kappa=0$ for
bosons and $\kappa=1$ for fermions, $c_0$ and $c_1$ are some
undetermined amplitudes, and the upper and lower signs correspond,
respectively, to the space-like and the time-like regions defined by
$x/2t>\sqrt{\beta}$ and $x/2t<\sqrt{\beta}$. Equation
(\ref{mainres}) also assumes the condition $|\Re \sqrt{\beta+i
\pi\kappa} -x/2t| >\Im \sqrt{ \beta+i \pi\kappa}$, where one should
take the positive branch of the square root, the meaning of which is
clarified in the main text. It might be argued that for any finite
$\kappa$, the second term in the parenthesis of the asymptotics
(\ref{mainres}) is exponentially small compared to the error term
and therefore should not appear there.  The presence of this term is
justified, however, by the fact that it becomes dominant in the
bosonic limit $\kappa\rightarrow 0$, when $\Im \lam_0^{\mp}
\rightarrow 0$. It is interesting to note that for all $\kappa\neq
0$, the first, leading term of the asymptotics (\ref{mainres}) is
not the one predicted by the conformal field theory or bosonization
\cite{CM,PKA}, and only the second, sub-leading term gives the
conformal part of the asymptotics, as demonstrated explicitly in
Section \ref{results}.

The plan of the paper is as follows. Section \ref{determinant}
describes the determinant representation for the correlation
functions obtained in \cite{PKA2}, which is used in Section
\ref{differentiale} to obtain differential equations indirectly
describing these functions. The relevant matrix Riemann-Hilbert
problem is introduced in Section \ref{Riemann}, and its asymptotic
solutions in the space-like and the time-like regions are  presented
in Sections \ref{RHPspace} and \ref{RHPtime}. The complete results
for the correlators are summarized in Section \ref{results}, and
their analysis in the bosonic, fermionic, and the low-temperature
(``conformal'') limit is given in Sections \ref{bflimit} and
\ref{CFT}. In the two Appendices we (A) discuss the large time and
distance asymptotic behavior of the correlators of free fermions,
and (B) present detailed analysis of the function
$C(x,t,\beta,\kappa)$.

\section{Determinant representation for the field-field
correlator}\label{determinant}

The second-quantized form of the Hamiltonian of the Lieb-Liniger gas
of anyons is
\be\label{hama} H=\int dx  \left( [\partial_x \fad(x)][\partial_x
\fa(x)]+c\fad(x)\fad(x)\fa(x)\fa(x)-h\fad(x)\fa(x) \right) , \ee
where $h$ is the chemical potential and  $c$ is the coupling constant, assumed in our case to be
infinite to make the anyons impenetrable. The anyonic fields satisfy
the commutation relations of the usual form
\[ \fa(x_1)\fad(x_2)=e^{-i\pi\kappa\epsilon(x_1-x_2)} \fad(x_2)
\fa(x_1)+ \delta(x_1-x_2)\, , \]
\[ \fad(x_1)\fad(x_2)=e^{i\pi\kappa\epsilon(x_1-x_2)} \fad(x_2)
\fad(x_1)\, , \]
with $\epsilon(x)=x/|x|,\ \epsilon(0)=0.$ The commutation relations
become bosonic for $\kappa=0$, and fermionic for $\kappa=1$. We are
interested in the asymptotic behavior of the space, time, and
temperature-dependent field-field correlator defined as
\[ \la\fa(x_2,t_2)\fad(x_1,t_1)\ra_T=\frac{\mbox{ Tr }(e^{-H/T}
\fa(x_2,t_2)\fad(x_1,t_1))}{\mbox{ Tr } e^{-H/T}}\, . \]
In \cite{PKA3}, we have obtained the following representation for
the correlator:
\be\label{correlator}
\la\fa(x_2,t_2)\fad(x_1,t_1)\ra_T=\left.e^{iht_{21}} \left(
\frac{1}{2\pi}G'(t_{12},x_{12})+\frac{\partial}{\partial\alpha}
\right) \det(1+\hat{ V}_T^\alpha)\right|_{\alpha=0}\, , \ee
where $x_{ab}=x_a-x_b,\ t_{ab}=t_a-t_b,\  a,b=1,2\, ,$ and
$\det(1+\hat{ V}_T)$ is the Fredholm determinant of the integral
operator with the kernel
\begin{eqnarray}\label{m1}
V_T^\alpha(\lam,\mu) &=&\cos^2(\pi\kappa/2)\exp\left\{-\frac{i}{2}t_{12}(
\lam^2+\mu^2) +\frac{i}{2}x_{12}(\lam+\mu)\right\} \sqrt{
\vartheta(\lam) \vartheta(\mu)}\nonumber\\ & &\ \ \times
\left[\frac{E(\lam|t_{12},x_{12})-E(\mu|t_{12},x_{12})}{\pi^2(\lam-\mu)}
-\frac{\alpha}{2\pi^3}E(\lam|t_{12},x_{12})E(\mu|t_{12},x_{12})\right]\,
,
\end{eqnarray}
which acts on an arbitrary function $f(\lam)$ as
\[ \left(V_T^\alpha f\right)(\lam)=\int_{-\infty}^\infty V_T^\alpha (
\lam,\mu) f(\mu) \ d\mu\, . \]
The functions $G'(t_{12},x_{12})$ and $E(\lam|t_{12},x_{12})$ in
Eqs.~(\ref{correlator}) and (\ref{m1}) are defined by
\[ G'(t_{12},x_{12})= \int_{-\infty}^\infty e^{it_{12}\mu^2-ix_{12}
\mu}\ d\mu\, , \]
and
\[ E(\lam|t_{12},x_{12})=\mbox{P.V.} \int_{-\infty}^\infty d\mu\
\frac{e^{it_{12}\mu^2-ix_{12}\mu}} {\mu-\lam}+ \pi\tan( \pi
\kappa/2)e^{it_{12}\lam^2-ix_{12}\lam}\, , \]
where $\mbox{ P.V.}$ denotes the Cauchy principal value, and
$\vartheta(\lam) \equiv \vartheta(\lam,T,h)$ in Eq.~(\ref{m1}) is
the Fermi distribution function of the quasiparticle momentum $\lam$
at temperature $T$ and chemical potential $h$:
\[ \vartheta(\lam,T,h)=\frac{1}{1+e^{(\lam^2-h)/T}}\, . \label{fw}
\]

The correlator (\ref{correlator}) depends on five variables: time,
distance, temperature, chemical potential and the statistics
parameter. It is convenient to rescale three of them and the
momentum $\lam$ by temperature:
\be\label{rescaled} x=(x_1-x_2)\sqrt{T}/2\, ,\ \ t=(t_2-t_1)T/2\, ,
\ \ \beta=h/T\, ,\ \ \lam\rightarrow\lam/\sqrt{T} \, . \ee
Then the explicit dependence of the correlator on temperature is
simple and is given by Eq.~(\ref{corr}). To see this, one needs first
to obtain a more manageable expression for the field correlator
(\ref{correlator}). In the rescaled variables (\ref{rescaled}), the
functions $G'$ and $E$ are given by
\be\label{defg} G'(t,x)=\sqrt{T}G(t,x)\, ,\ \ \ G(t,x)=
\int_{-\infty}^\infty e^{-2it\lam^2-2ix\lam}\ d\lam\, , \ee
and
\be\label{defe}  E(\lam|t,x)=\mbox{P.V.} \int_{-\infty}^\infty d\mu\
\frac{e^{ -2it\mu^2-2ix\mu}}{\mu-\lam}+ \pi\tan( \pi
\kappa/2)e^{-2it\lam^2- 2ix\lam}\, . \ee
We introduce the two functions $e_\pm(\lam)$:
\be\label{em}
e_-(\lam)=\frac{\cos(\pi\kappa/2)}{\pi}\sqrt{\vartheta(\lam)}
e^{it\lam^2+i\lam x}\, , \ee
and
\be \label{ep} e_+(\lam)=e_-(\lam)E(\lam)= \frac{ \cos(
\pi\kappa/2)}{\pi}\sqrt{\vartheta(\lam)}e^{it\lam^2+i\lam x}
\left(\mbox{P.V.} \int_{-\infty}^\infty d\mu\ \frac{
e^{-2it\mu^2-2ix\mu}}{\mu-\lam}+ \pi\tan( \pi
\kappa/2)e^{-2it\lam^2-2ix\lam}\right) \, . \ee
In terms of these functions, the kernel (\ref{m1}) of the integral
operator appearing in (\ref{correlator}) is expressed as
\[ V_T^\alpha(\lam,\mu)=V_T(\lam,\mu)-\frac{\alpha}{2\pi}A_T(\lam,
\mu)\, , \]
with
\be\label{factorization}
V_T(\lam,\mu)=\frac{e_+(\lam)e_-(\mu)-e_-(\lam)e_+(\mu)}{\lam-\mu}\, ,
\ee
and
\[  A_T(\lam,\mu)=e_+(\lam)e_+(\mu)\, . \]
In what follows, we also need some basic formulae from the theory of
Fredholm determinants:
\[ \log\det(1+\hat V)=\sum_{n=1}^\infty\frac{(-1)^{n+1}}{n} \mbox{Tr }
V^n\, ,\ \ \  (1+\hat V)^{-1}=1-V^1+V^2+ ... \, , \]
where $V^n(\lam,\mu)$ is determined successively by
$V^n(\lam,\mu)=\int V(\lam,\nu)V^{n-1}(\nu,\mu)d\nu$ and
$V^1(\lam,\mu)=V(\lam,\mu)$. The trace is defined naturally as
$\mbox{Tr } V=\int V(\lam,\lam)d\lam$. Then, $\mbox{Tr }
V^2=\int\int V(\lam,\mu)V(\mu,\lam)d\lam d\mu$, and so on. Using
these relations, one can see directly that
\[ \left.\frac{\6\det(1+V_T^\alpha)}{\6 \alpha}\right|_{\alpha=0}=
-\frac{\sqrt{T}}{2\pi}\mbox{Tr }[(1+\hat V_T)^{-1}\hat A_T]\det(1+
\hat V_T)\, , \]
which together with Eq.~(\ref{defg}) gives Eq.~(\ref{corr}) for the
correlator with
\be\label{m2}
g(x,t,\beta,\kappa)=-\frac{1}{2\pi}e^{2it\beta}\left(\mbox{Tr }[
(1+\hat V_T)^{-1}\hat A_T]-G(t,x)\right)\det(1+\hat V_T)\, . \ee

The integral operator $\hat V_T$ whose determinant appears in
(\ref{m2}) is of a special type called ``integrable'' operators
\cite{IIKS,KBI,HI}. This type of integral operators have kernels of
the ``factorizable'' structure similar to Eq.~(\ref{factorization}),
and are ubiquitous in investigations of correlations functions of
integrable quantum systems and distribution of eigenvalues of random
matrices. If an operator is integrable, the resolvent operator
defined as
\[ \hat R_T=(1+\hat V_T)^{-1}\hat V_T\, ,\ \ \ (1+\hat V_T)(1-\hat
R_T)=1\, , \]
is also of the same type, which means that the resolvent kernel that
solves the integral equation
\[ R_T(\lam,\mu)+\int_{-\infty}^{+\infty}V_T(\lam,\nu)R_T(\nu,\mu)
\ d\nu=V_T(\lam,\mu)\, , \]
is also factorized as in (\ref{factorization}):
\[ R_T=\frac{f_+(\lam)f_-(\mu)-f_-(\lam)f_+(\mu)}{\lam-\mu}\, . \]
The functions $f_\pm(\lam)$ are the solutions of the
integral equations
\be\label{deff} f_\pm(\lam)+\inti V_T(\lam,\mu)f_\pm(\mu)
d\mu=e_\pm(\lam)\, . \ee
Now we can introduce an important class of objects
called auxiliary potentials  defined as
\be\label{defb} B_{lm}(x,t,\beta,\kappa)=\inti e_l(\lam) f_m(\lam)
d\lam\, , \ \ l,m=\pm\, , \ee
and
\be\label{defc} C_{lm}(x,t,\beta,\kappa)=\inti \lam e_l(\lam)
f_m(\lam) d\lam\, , \ \ l,m=\pm\, . \ee
Due to the symmetry of $V_T(\lam,\mu)$, we have $B_{+-}=B_{-+}$. One
can see directly that $\mbox{Tr}[(1+\hat V_T)^{-1}\hat A_T]=B_{++}$.
Therefore, defining $b_{++}=B_{++}-G$, we obtain the following
representation for the function (\ref{m2}) in the time- and
temperature-dependent correlator (\ref{corr}):
\be\label{deffcpot} g(x,t,\beta,\kappa)=
-\frac{1}{2\pi}e^{2it\beta}b_{++}(x,t,\beta,\kappa)\det(1+\hat
V_T)\, . \ee
Since $g(x,t,\beta,\kappa)=g(-x,t,\beta,-\kappa)$ and $g(x,t,\beta,
\kappa)=g^*(x,-t,\beta,-\kappa)$, to study the correlator, it is
sufficient to investigate only the case $x>0,\ t>0$.

\section{Differential equations for the correlation  functions}
\label{differentiale}

Obtaining the differential equations directly for the correlation
functions at finite temperature is an extremely difficult task. One
can, however, obtain a system of partial differential equations for
the auxiliary potentials, and show that the derivatives of the
logarithm of the Fredholm determinant
\be
\sigma(x,t,\beta,\kappa)=\log\det(1+\hat V_T)\, ,
\ee
is expressed in terms of a combination of the auxiliary potentials
and derivatives. The differential equation for the potentials are
obtained as follows. First, we define a two-component function
\[ F(\lam)=\left(\begin{array}{c} f_+(\lam)\\f_-(\lam)\end{array}
\right), \]
and look for three matrix operators $\textsf{L}(\lam),\
\textsf{M}(\lam),\  \textsf{N}(\lam)$ which depend on the auxiliary
potentials and their derivatives and satisfy the Lax representation
conditions
\[ \textsf{L}(\lam)F(\lam)=0\, , \ \ \textsf{M}(\lam)F(\lam)=0\, ,\
\ \  \textsf{N}(\lam)F(\lam)=0\, . \]
The differential equations for the potentials are obtained then from
the compatibility conditions for the Lax representation
\[ [\textsf{L}(\lam),\ \textsf{M}(\lam)]=[\textsf{L}(\lam),\
\textsf{N}(\lam)]=[\textsf{M}(\lam),\ \textsf{N}(\lam)]=0 \]
which should be valid for any value of the spectral parameter $\lam$.

Specific calculations follow closely those for the impenetrable
bosons \cite{IIKS,KBI}, and their main ingredient are the following
relations
\bea
\6_xE(\lam)&=&-2iG-2i\lam E(\lam)\, ,\nonumber\\
\6_tE(\lam)&=&-2i\lam^2E-2i\lam G+\6_xG\, ,\nonumber\\
\6_\beta E(\lam)&=&0\, ,\\
\6_\lam E(\lam)&=&-(4it\lam+2ix)E-4itG\, ,\nonumber\\
\eea
which can be proved directly from the definitions (\ref{defg}) and
(\ref{defe}) of the functions $G$ and $E$. Here we only present the
results of the calculations.

\begin{lem}
The potentials $C(x,t,\beta,\kappa)$ can be expressed in terms of
the potentials $B(x,t,\beta,\kappa)$ and $G(x,t)$ as follows
\[ C_{++}=\frac{i}{2}\6_xB_{++}-2GB_{+-}+B_{+-}B_{++}\, , \]
\[ C_{--}=-\frac{i}{2}\6_x B_{--}-B_{+-}B_{--}\, , \]
and
\[ C_{+-}=C_{-+}=B_{+-}^2-B_{++}B_{--}\, . \]
\end{lem}
\begin{thm}
Define
\[ g_-\equiv e^{-2it\beta}B_{--},\ \ \  g_+\equiv e^{2it\beta}b_{++}\,
, \]
and
\[n\equiv g_-g_+=b_{++}B_{--},\ \ \  p\equiv g_-\6_xg_+-g_+\6_xg_-\,
. \]
Then $g_-$ and $g_+$ satisfy the separated nonlinear Schr\"odinger equation
\bea\label{NSSE}
-i\6_tg_+&=&2\beta g_++\frac{1}{2}\6_x^2g_++4g_+^2g_-\, ,\nonumber\\
i\6_tg_-&=&2\beta g_-+\frac{1}{2}\6_x^2g_-+4g_-^2g_+\, ,
\eea
and
\[ -2i\6_t n=\6_x p\, . \]
The equations containing the $\beta$-derivatives are
\[ \frac{\6_\beta\6_x g_+}{g_+}=\frac{\6_\beta\6_x g_-}{g_-}=\varphi\,
, \]
and
\[ -i\6_t\varphi+4\6_\beta p=0\, , \ \ \ \6_x\varphi+8\6_\beta n+2=0\,
. \]
\end{thm}

The previous theorem characterizes completely the potentials
$B_{--}$ and $b_{++} (B_{++})$. The other potentials can be
expressed in terms of these two as
\be\label{bpmx}
\6_xB_{+-}=2ib_{++}B_{--}\, ,\ \ \ \6_t B_{+-}=-p\, ,
\ee
\[ \6_\beta B_{+-}=-ix/4-i\varphi/4\, , \]
and
\be\label{e37}
\6_x(C_{+-}-C_{-+})=(B_{++}-2G)\6_x B_{--}-B_{--}\6_x B_{++}\, .
\ee
Finally, we have the following theorem
\begin{thm}
The derivatives of the logarithm of the Fredholm determinant
$\sigma(x,t,\beta,\kappa)$ are given by
\bea\label{desigma} \6_x\sigma&=&-2iB_{+-}\, ,\nonumber\\
\6_t\sigma&=&-2iGB_{--} -2i(C_{+-}+C_{-+})\, ,\\
\6_\beta\sigma&=&-2it\6_\beta(C_{+-}+C_{-+})-2ix \6_\beta
B_{+-}-2itB_{--}\6_\beta B_{++}+2it(B_{++}-2G) \6_\beta
B_{--}\nonumber\\ & &\ \ \ \ \ \ \ +2(\6_\beta B_{++})(\6_\beta
B_{--})-2(\6_\beta B_{+-})^2\, .\nonumber \eea
\end{thm}
All the differential equations above do not depend on the statistics
parameter, and are the same as those obtained for impenetrable
bosons in \cite{IIKS,KBI}. The statistics parameter appears only in
the initial conditions which can be extracted from the equal-time
field correlator studied in \cite{PKA3,PKA4}. The same phenomenon
was noticed also for static correlators at $T=0$ \cite{SC} and
finite temperature \cite{PKA3,PKA4}.

\section{Matrix Riemann-Hilbert Problem}\label{Riemann}

The discussion in the previous sections implies that with the use of
the differential equations, the large-time and -distance asymptotic
behavior of the field correlator can be extracted from the
corresponding behavior of the auxiliary potentials. A powerful
method of obtaining the asymptotics for the potentials is the
formalism of the matrix Riemann-Hilbert problem (RHP). Here, we
consider a specific matrix RHP associated with the integrable system
that characterizes the potentials. Solution of this RHP will allow
us to obtain the asymptotics of the potentials and field correlator.
In details, we are interested in finding a $2\times 2$ matrix
function $\chi(\lam)$, nonsingular for all $\lam\in \mathbb{C},$ and
analytic separately in the upper and lower half planes, which also
satisfies the following conditions
\bea\label{RHPT} \chi_-(\lam)&=&\chi_+(\lam)G(\lam)\, ,\ \ \
\chi_\pm(\lam)= \lim_{\epsilon\rightarrow 0^+}\chi(\lam\pm
i\epsilon)\, ,\ \ \lam\in\mathbb{R}\, ,\nonumber\\
\chi(\infty)&=&I\, . \eea
Here $I$ is the unit $2\times 2$ matrix and $G(\lam)$ is the
conjugation matrix defined only for real $\lam$ and given in our
case by
\be G(\lam)=\left(\begin{array}{lr} 1-2\pi i e_+(\lam)e_-(\lam) &
2\pi i e_+^2(\lam)\\ -2 \pi i e_-^2(\lam) & 1+2\pi i
e_+(\lam)e_-(\lam) \end{array}\right) \ee
The functions $e_\pm(\lam)$ appearing in this equation are defined
in (\ref{ep}) and (\ref{em}). The matrix function $\chi(\lam)$
depends also on $x,t,\beta,$ and $\kappa$, but this dependence is
suppressed in our notations. Also, in what follows we will consider
$\kappa\in(0,1]$. The case of impenetrable bosons, $\kappa=0$,
requires a special treatment presented in \cite{IIKV,KBI}.

\subsection{Connection with the auxiliary potentials}\label{conaux}

In this section, we show that in the limit of large $\lam$, the
auxiliary potentials can be extracted from the solution of the RHP
(\ref{RHPT}). To do this, one can see first (as shown, e.g., in Chap.
XV of \cite{KBI}) that the RHP is equivalent to the following system
of singular integral equations
\[ \chi_+(\lam)=I+\frac{1}{2\pi i}\inti \frac{\chi_+(\mu)[I-G(\mu)]
}{\mu-\lam-i0}d\mu\, ,\ \ \lam \in\mathbb{R}\, . \]
Multiplying from the right with
\[ H(\lam)=\left(\begin{array}{lr} 1&e_+(\lam)\\
                    0&e_-(\lam)
               \end{array}\right), \]
and introducing $\hat\chi(\lam)=\chi_+(\lam)H(\lam)$, we transform
these equations into
\begin{eqnarray*}
\hat\chi(\lam)&=&H(\lam)+\frac{1}{2\pi i}\inti\frac{\chi_+(\mu)H(\mu)
H^{-1}(\mu)[I-G(\mu)]H(\lam)}{\mu-\lam-i0}d\mu\, ,\nonumber\\
&=&H(\lam)+\frac{1}{2\pi i}\inti\frac{\hat\chi(\mu)\hat G(\lam,\mu)
}{\mu-\lam-i0}d\mu\, , \end{eqnarray*}
where
\[ \hat G(\lam)=\left(\begin{array}{cc} 0 & 0\\
2 \pi i e_-(\mu) & 2\pi i(e_+(\lam)e_-(\mu)-e_-(\lam)e_+(\mu) )
\end{array}\right) , \]
and
\be\label{m4} \hat \chi(\lam)=\chi_+(\lam)H(\lam)
=\left(\begin{array}{lr} \chi_{11,+}(\lam) &
\chi_{11,+}(\lam)e_+(\lam)+\chi_{12,+}(\lam) e_-(\lam)\\
\chi_{21,+}(\lam) &
\chi_{21,+}(\lam)e_+(\lam)+\chi_{22,+}(\lam)e_-(\lam)
\end{array}\right) . \ee
The integral equations for $\hat\chi_{12}$ and $\hat\chi_{22}$ are
\be\label{m5}
\hat\chi_{12}=e_+(\lam)+\inti\frac{\hat\chi_{12}(\mu)(e_+(\lam)e_-(\mu)
-e_-(\lam)e_+(\mu))}{\mu-\lam}\ d\mu \, , \ \ \lam\in\mathbb{R}\, ,
\ee
and
\be\label{m6}
\hat\chi_{22}=e_-(\lam)+\inti\frac{\hat\chi_{22}(\mu)(e_+(\lam)e_-(\mu)
-e_-(\lam)e_+(\mu))}{\mu-\lam}\ d\mu \, , \ \ \lam\in\mathbb{R}\, .
\ee
Taking into account that the functions $f_\pm(\lam)$ satisfy the
integral equations (\ref{deff}), where the kernel $V_T(\lam,\mu)$ is
given by Eq.~(\ref{factorization}), one can see directly from
(\ref{m5}) and (\ref{m6}) that
\be\label{m7}
\hat \chi_{12}(\lam)=f_+(\lam)\, ,\ \ \ \hat\chi_{22}(\lam)=f_-(\lam)\, .
\ee
Also, Eq.~(\ref{m4}) gives that
$\hat\chi_{11}(\lam)=\chi_{11,+}(\lam)\, ,\
\hat\chi_{21}(\lam)=\chi_{21,+}(\lam)$. Therefore, using
Eq.~(\ref{m7}) we obtain the following integral equations for
$\chi_{11,+}(\lam)$ and $\chi_{21,+}(\lam)$
\[ \chi_{11,+}(\lam)=1+\inti\frac{e_-(\mu)f_+(\mu)}{\mu-\lam-i0}\
d\mu\, ,\ \ \ \chi_{21,+}(\lam)=\inti\frac{e_-(\mu)f_-(\mu)
}{\mu-\lam-i0} \ d\mu\, , \;\;\; \lam\in\mathbb{R}. \]
Continuing analytically into the upper half plane, taking the limit
of large $\lam$, and using the definitions of the auxiliary
potentials (\ref{defb}) and (\ref{defc}), we obtain from these
equations
\be\label{m8}
\chi_{11}(\lam)=1-\frac{1}{\lam}B_{-+}-\frac{1}{\lam^2}C_{-+}+
O\left(\frac{1}{\lam^3}\right), \ee \be\label{m9}
\chi_{21}(\lam)=-\frac{1}{\lam}B_{--}-\frac{1}{\lam^2}C_{--}+
O\left(\frac{1}{\lam^3}\right). \ee
In order to obtain similar expansions for $\chi_{12}(\lam)$ and
$\chi_{22}(\lam)$, we proceed in the following fashion. First,
Eq.~(\ref{m4}) gives:
\[ \hat\chi_{22}(\lam)=f_-(\lam)=\chi_{21,+}(\lam)e_+(\lam)+
\chi_{22,+}(\lam)e_-(\lam)\, . \]
Then, using Eq.~(\ref{m9}) and the large-$\lam$ expansion of the
integral equation (\ref{deff}) defining $f_-(\lam)$ we rewrite this
equation as
\begin{equation*}
e_-(\lam)-\frac{1}{\lam}\left(e_+(\lam)B_{--}-e_-(\lam)B_{-+}\right)
-\frac{1}{\lam^2}(e_+(\lam)C_{--}-e_-(\lam)C_{+-})+\cdots =
\end{equation*}
\be \ \ \ \ \ \ \ \ \ =e_+(\lam)\left(-\frac{1}{\lam}B_{--}
-\frac{1}{\lam^2}C_{--}\right)+ e_-(\lam)\left(1+\frac{1}{\lam}
\chi_{22}^{(1)}+\frac{1}{\lam^2}\chi_{22}^{(2)}\right)+\cdots\ \ee
Comparison of the two sides of this equation implies that
\be\label{m10}
\chi_{22}(\lam)=1+\frac{1}{\lam}B_{+-}-\frac{1}{\lam^2}C_{+-}+
O\left(\frac{1}{\lam^3}\right) . \ee
The expansion for $\chi_{12}(\lam)$ can be derived through similar
steps:
\be\label{m11}
\chi_{12}(\lam)=\frac{1}{\lam}B_{++}+\frac{1}{\lam^2}C_{-+}+
O\left(\frac{1}{\lam^3}\right) . \ee
Collecting the results (\ref{m8}), (\ref{m9}), (\ref{m10}) and
(\ref{m11}), we see that in the large-$\lam$ limit, the auxiliary
potentials follow from the expansion of the solution of the RHP
(\ref{RHPT}):
\[ \chi(\lam)=I+\frac{1}{\lam}\left(\begin{array}{lr} -B_{-+}& B_{++}\\
-B_{--}&B_{+-} \end{array}\right) +\frac{1}{\lam^2}
\left(\begin{array}{lr} -C_{-+}& C_{++}\\ -C_{--}&C_{+-}
\end{array}\right)+O\left(\frac{1}{\lam^3}\right) , \ \
\lam\rightarrow \infty\, . \]
%

\subsection{Transformations of the RHP}

It will be useful to perform several transformations on the RHP
(\ref{RHPT}). The first one is
\[ \chi(\lam)=\tilde\chi(\lam)\chi_0(\lam)\, , \]
with
\[ \chi_0(\lam)=\left(\begin{array}{cc} 1& -a(\lam)\\
0&1 \end{array}\right)\, ,\ \ \
a(\lam)=\inti\frac{e^{-2it\mu^2-2ix\mu}}{\mu-\lam}\ d\mu\, . \]
Using the fact that the boundary values of the function $a(\lam)$ on
the real axis are:
\[ a_\pm(\lam)=\pm i \pi e^{-2it\lam^2-2ix\lam}+\mbox{P.V.}\inti
\frac{e^{-2it\mu^2-2ix\mu}}{\mu-\lam}\ d\mu\, , \]
it can be shown that the matrix $\tilde\chi(\lam)$ solves the
transformed RHP
\be \tilde\chi_-(\lam)=\tilde\chi_+(\lam)\tilde G(\lam)\, ,\ \ \
\lam\in\mathbb{R}\, ;\ \ \ \tilde\chi(\infty)=I\, ,  \ee
with $\tilde G(\lam)=\chi_{0+}(\lam)G(\lam)\chi_{0-}^{-1}(\lam)$
given explicitly by
\[ \tilde G(\lam)=\left(\begin{array}{lr} 1-\vartheta(\lam)(1+
e^{i\pi\kappa})& 2\pi i(\vartheta(\lam)-1)e^{-2it\lam^2-2i\lam x}\\
-\frac{2i}{\pi}\cos^2(\pi\kappa/2)\vartheta(\lam)e^{2it\lam^2+2i\lam
x}&1-\vartheta(\lam)(1+e^{-i\pi\kappa})
\end{array}\right)  . \]

The specific form of the second transformation depends on whether we
are considering the ``space-like'' $(x/2t>\sqrt{\beta})$ or the
``time-like'' $(x/2t<\sqrt{\beta})$ region.

\subsubsection{Transformation in the space-like case}

As a first step, we need to introduce the functions
\be
\varphi(\lam^2,\beta,\kappa)=\frac{e^{\lam^2-\beta}-e^{i\pi\kappa}
}{e^{\lam^2-\beta}+1}\, , \ee
and
\be \alpha(\lam)=\exp\left\{-\frac{1}{2\pi i}\inti
\frac{d\mu}{\mu-\lam} \ln \varphi(\mu^2,\beta,\kappa)\right\}\, .
\ee
The latter is the solution of the following scalar Riemann-Hilbert
problem (for more information on scalar RHP see, e.g., \cite{G})
\begin{equation*}
\alpha_-(\lam)=\alpha_+(\lam)[1-\vartheta(\lam)(1+e^{i\pi\kappa})]\,
, \ \ \lam\in\mathbb{R}\, ; \ \ \ \alpha(\infty)=1\, .
\end{equation*}
Then, the second transformation in the space-like case is:
\[ \Phi(\lam)=\tilde\chi(\lam)e^{-\sigma_3\ln \alpha(\lam)}\, , \]
where $\sigma_3$ is the third Pauli matrix. The new matrix function
$\Phi(\lam)$ solves the matrix RHP
\be\label{RHPT2} \Phi_-(\lam)=\Phi_+(\lam) G_\Phi(\lam)\, ,\ \ \
\lam\in\mathbb{R}\, ; \;\;\; \tilde\chi(\infty)=I\, , \ee
with the conjugation matrix $G_\Phi(\lam)=e^{\sigma_3\ln
\alpha(\lam)}\tilde G(\lam)e^{-\sigma_3\ln \alpha(\lam)}$:
\be\label{m12}
G_\Phi(\lam)=\left(\begin{array}{lr} 1&p(\lam) e^{-2it\lam^2-2ix\lam}\\
q(\lam) e^{2it\lam^2+2ix\lam}& 1+p(\lam)q(\lam) \end{array}\right)\,
, \ee
where
\be\label{defpspace} p(\lam)=-2\pi i[\alpha_-(\lam)]^2
\frac{e^{\lam^2-\beta}}{ e^{\lam^2-\beta}-e^{i\pi\kappa}}\, , \ee
and
\be\label{defqspace}
q(\lam)=-\frac{2i}{\pi}\cos^2(\pi\kappa/2)[\alpha_+(\lam)]^{-2}
\frac{1}{e^{\lam^2-\beta}-e^{i\pi\kappa}}\, .
\ee

\subsubsection{Transformation in the time-like case}

The transformation in the time-like case is similar to the one
performed in the space like case. The difference is that the
function $\alpha(\lam)$ is now defined as (note the change of the
sign of $\kappa$)
\[ \alpha(\lam)=\exp\left\{-\frac{1}{2\pi i}\inti \frac{d\mu}{\mu-\lam}
\ln \varphi(\mu^2,\beta,-\kappa)\right\} , \]
and is the solution of the scalar Riemann-Hilbert problem
\begin{equation*}
\alpha_-(\lam)=\alpha_+(\lam)[1-\vartheta(\lam)(1+e^{-i\pi\kappa})]\,
, \ \ \lam\in\mathbb{R}\, ; \ \ \ \alpha(\infty)=1\, .
\end{equation*}
The new matrix $\Phi(\lam)=\tilde\chi(\lam)e^{+\sigma_3\ln
\alpha(\lam)}$ solves the same RHP  (\ref{RHPT2}) but now with the
conjugation matrix $G_\Phi(\lam)=e^{-\sigma_3\ln \alpha(\lam)}\tilde
G(\lam)e^{+\sigma_3\ln \alpha(\lam)}$:
\be\label{g1} G_\Phi(\lam)=\left(\begin{array}{lr} 1+p(\lam)q(\lam)
&p(\lam) e^{-2it\lam^2-2ix\lam}\\ q(\lam) e^{2it\lam^2+2ix\lam}&1
\end{array}\right)\, , \ee
where $p(\lam)$ and $q(\lam)$ are
\be\label{defptime} p(\lam)=-2\pi i[\alpha_-(\lam)]^2 \frac{e^{\lam^2
-\beta}}{e^{\lam^2 -\beta}-e^{-i\pi\kappa}}\, , \ee
and
\be\label{defqtime}
q(\lam)=-\frac{2i}{\pi}\cos^2(\pi\kappa/2)[\alpha_+(\lam)]^{-2}
\frac{1}{e^{\lam^2-\beta}-e^{-i\pi\kappa}}\, .
\ee
%

\subsection{Potentials in terms of the $\Phi$ matrix}\label{potentials}

In Section \ref{conaux}, we showed that the auxiliary potentials we
can be extracted from the large-$\lam$ expansion of the solution
$\chi(\lam)$ of the RHP (\ref{RHPT}). However, since we explicitly
will be finding the asymptotic solution of the RHP (\ref{RHPT2}), we
need to express the potentials in terms of the $\Phi$ matrix. The
computations necessary to do this are presented below only in the
space-like case, the time-like case being similar. The first step is
to obtain the large-$\lam$ expansion of all the terms in the
relation
\[ \Phi(\lam)=\chi(\lam)\chi_0^{-1}(\lam)e^{-\sigma_3\ln
\alpha(\lam)}\, . \]
Explicitly, we have in the limit $\lam\rightarrow \infty$:
\[ \chi(\lam)=I+\frac{1}{\lam}\left(\begin{array}{lr} -B_{-+}&
B_{++}\\ -B_{--}&B_{+-} \end{array}\right) +\frac{1}{\lam^2}
\left(\begin{array}{lr} -C_{-+}& C_{++}\\
-C_{--}&C_{+-} \end{array}\right)+O\left(\frac{1}{\lam^3}\right) ,
\]
\[ \chi_0^{-1}(\lam)=I+\frac{1}{\lam}\left(\begin{array}{cc} 0& -G\\
0&0 \end{array}\right)+\frac{1}{\lam}\left(\begin{array}{cc}0 &
-G^{(1)}\\ 0&0 \end{array}\right)+O\left(\frac{1}{\lam^3}\right), \]
\[ e^{-\sigma_3\ln\alpha(\lam)}=I+\frac{1}{\lam}\left(\begin{array}{cc}
-\alpha_0& 0\\ 0&\alpha_0 \end{array}\right)
+O\left(\frac{1}{\lam^3}\right) , \]
where $G$ is given by (\ref{defg}), and
\be\label{defalpha0}
\alpha_0=\frac{1}{2\pi i}\inti\ln \varphi(\mu^2,\beta,\kappa)\  d\mu \, .
\ee
Considering a similar expansion for $\Phi(\lam)$
\[ \Phi(\lam)=I+\frac{1}{\lam}\left(\begin{array}{lr} (\Phi_1)_{11}&
(\Phi_1)_{12}\\ (\Phi_1)_{21}& (\Phi_1)_{22} \end{array}\right)
+\frac{1}{\lam^2}\left(\begin{array}{lr} (\Phi_2)_{11}& (\Phi_2)_{12}\\
(\Phi_2)_{21}& (\Phi_2)_{22} \end{array}\right)
+O\left(\frac{1}{\lam^3}\right)\, ,\ \ \lam\rightarrow \infty\, , \]
and equating the terms with equal powers of $\lam$, one finds
\bea\label{potsl} B_{+-}=-(\Phi_1)_{11}-\alpha_0\, , \ \
b_{++}&=&(\Phi_1)_{12}\, , \ \ B_{--}=-(\Phi_1)_{21}\, ,\nonumber\\
C_{+-}+C_{-+}+B_{--}G&=&(\Phi_2)_{22}-(\Phi_2)_{11}\, . \eea
In the time-like case, similar computations give
\bea \label{pottl} B_{+-}=-(\Phi_1)_{11}+\alpha_0\, ,\ \ b_{++}&=&(\Phi_1)_{12}\, ,\ \ B_{--}=-(\Phi_1)_{21}\, ,\nonumber\\
C_{+-}+C_{-+}+B_{--}G&=&(\Phi_2)_{22}-(\Phi_2)_{11}\, , \eea
with
\be\label{defalpha0t}
\alpha_0=\frac{1}{2\pi i}\inti  \ln \varphi(\mu^2,\beta,-\kappa)\ d\mu\, .
\ee
%

\section{Asymptotic solution of the RHP. Space-like case}\label{RHPspace}

We are interested in solving the RHP (\ref{RHPT2}) in the limit of
large $x>0$ and $t>0$, but with finite ratio $x/t=const$. If one
compares the solution with the corresponding solution for the static
case (without time $t$), the analysis in the time-dependent case is
more complicated due to the presence of the stationary point of the
phase
\[ \thet(x,t,\lam)\equiv t\lam^2+x\lam\, , \]
where $\6_\lam\thet=0$. This condition gives
\[ \lam_s=-\frac{x}{2t}\, , \; x>0\, ,\; t>0\, ,\; \lam_s<0\, . \]
The asymptotic analysis of the RHP has to properly take into account
this stationary point. In this work, we do this by employing the
method pioneered in \cite{M,I}, also used for the impenetrable
bosons \cite{IIKV,KBI}. The main ingredient of this approach is the
Manakov ansatz \cite{M} which provides an approximate solution
$\Phi^m(\lam)$ to the RHP (\ref{RHPT2}). The Manakov ansatz in the
space-like region is different from the one in the time-like region,
even though the results obtained from both forms of ansatz will be
the same in the leading order. The asymptotic analysis is based on
the two assumptions: (i) the RHP is solvable, and (ii) the boundary
values of $\Phi_\pm(\lam)$ on the real axis are uniformly bounded in
the limit $t\rightarrow \infty$. These assumptions can be proved
following the Sections 6 and 7 of \cite{IIKV}. Also, we require that
\be\label{c100} \left | |\Re \sqrt{\beta+i \pi\kappa}| -x/2t\right|
>|\Im \sqrt{\beta+i \pi\kappa}| \, . \ee
The meaning of this inequality is discussed below (see Sec.~\ref{abp}). While this condition  is not
 essential in that one can analyze other regimes as well, it is always satisfied, in particular, in the more interesting low-temperature case $\beta >>1$.

\subsection{Manakov ansatz}
The space-like region is defined by
\[ \lam_s<-\sqrt{\beta}\, ,\ \beta=h/T>0\, , \]
condition that can be expressed in more explicit notations as
\[ (x_1-x_2)>v_F(t_2-t_1)>0\, , \]
where $v_F$ is the velocity of excitations, which in our model of
impenetrable anyons coincides with the Fermi velocity of free
fermions, $v_F=2 k_F$, with $k_F= \sqrt{h}$, in the conventions used
in the Hamiltonian (\ref{hama}). The Manakov ansatz in the
space-like region is given by
\[ \Phi^m(\lam)=\left(\begin{array}{cc} 1&-I^p(\lam)\\
-I^q(\lam)&1 \end{array}\right)e^{\sigma_3\ln\delta(\lam)} \, , \]
where
\be I^p(\lam)=\frac{1}{2\pi i}\inti \frac{\delta_+(\mu)\delta_-( \mu)
}{\mu-\lam}p(\mu)e^{-2i\thet(x,t,\mu)}d\mu\, , \label{aip} \ee
\be I^q(\lam)=\frac{1}{2\pi i}\inti \frac{\delta_+^{-1}(\mu)
\delta_-^{-1} (\mu)}{\mu-\lam}q(\mu)e^{2i\thet(x,t,\mu)}d\mu\, ,
\label{aiq} \ee
and the functions $p(\lam)$ and $q(\lam)$ defined by
Eqs.~(\ref{defpspace}) and (\ref{defqspace}). The function
$\delta(\lam)$ is the solution of the following scalar RHP
\begin{equation*}
\delta_+(\lam)=\delta_-(\lam)[1+p(\lam)q(\lam)\eta(\lam_s-\lam)] \,
,\ \ \lam\in\mathbb{R}\, ,\;\;\; \delta(\infty)=1\, ,
\end{equation*}
with $\eta(\lam)$ denoting the step function
\[ \eta(\lam)=\left\{\begin{array}{c} 1\, , \;\; \lam>0\, ,\\ 0\, ,
\;\; \lam<0\, .\\ \end{array}\right. \]
This scalar RHP problem can be solved explicitly (see, e.g.,
\cite{G}), and if we take into account that
\[ 1+p(\lam)q(\lam)=|\varphi(\lam^2,\beta,\kappa)|^2\, , \]
the solution is
\[ \delta(\lam)=\exp\left\{\frac{1}{2\pi i}\int_{-\infty}^{\lam_s}
\frac{d\mu}{\mu-\lam}\ln|\varphi(\mu^2,\beta,\kappa)|^2\right\}\, .
\]

\subsubsection{Properties of\ $\delta(\lam)$}

Before we show that $\Phi^m(\lam)$ is an approximate solution of the
RHP (\ref{RHPT2}), it is useful to investigate some of the
properties of the function $\delta(\lam)$. For $\lam\in(\lam_s,
\infty)$, integration by parts gives
\[ \frac{1}{2\pi i}\int_{-\infty}^{\lam_s}\frac{d\mu}{\mu-\lam}\ln
|\varphi(\mu^2,\beta,\kappa)|^2= \frac{1}{\pi i}\ln(\lam-\lam_s)
\ln|\varphi(\lam_s^2,\beta,\kappa)|- \frac{1}{\pi i} \int_{-\infty
}^{\lam_s} \ln|\mu-\lam|d (\ln|\varphi(\mu^2,\beta,\kappa)|)d\mu\, .
\]
Introducing two quantities:
\be\label{defnu}
\nu(\lam_s,\beta,\kappa)=\frac{1}{\pi}\ln|\varphi(\lam_s^2,\beta,
\kappa)|^{-1}>0\, , \ee
and
\[\gamma(\lam)=\frac{1}{\pi}\int_{-\infty}^{\lam_s}\ln|\mu-\lam|
d (\ln|\varphi(\mu^2,\beta,\kappa)|)d\mu\, , \]
one can use this relation to write $\delta_{\pm}(\lam)$ for
$\lam\in(\lam_s,\infty)$ as
\[ \delta_{\pm}(\lam)=(\lam-\lam_s)_\pm^{i\nu}\exp(i\gamma(\lam))\,
, \]
where $(\lam-\lam_s)_\pm^{i\nu}$ are the boundary values of the
multi-valued function $(\lam-\lam_s)^{i\nu}$ defined in the complex
plane with the branch cut along the ray $(-\infty,\lam_s]$. When
$\lam\in(-\infty,\lam_s)$, integration by parts for
singular integrals (see, e.g., \cite{G}, pp.~18) gives
\[ \delta_{\pm}(\lam)=\exp\left\{\pm \ln|\varphi(\lam^2,\beta,\kappa)|
+\frac{1}{\pi i}\ln(\lam_s-\lam)_{\pm}\ln|\varphi(\lam_s^2,\beta,
\kappa)| -\frac{1}{\pi i}\int_{-\infty}^{\lam_s}\ln|\mu-\lam|d(\ln|
\varphi(\mu^2,\beta,\kappa)|)d\mu \right\} . \]
This equation can be rewritten as
\[ \delta_{\pm}(\lam)=(\lam-\lam_s)_{\pm}^{i\nu}\exp(i\gamma(\lam))
|\varphi (\lam^2,\beta,\kappa)|^{\pm 1}|\varphi(\lam_s^2,\beta,
\kappa)|^{\mp 1}\, . \]
in the notations used above. Therefore, the function $\delta(\lam)$
in both regions is
\[ \delta_{\pm}(\lam)=(\lam-\lam_s)_{\pm}^{i\nu}\exp(i\gamma(\lam))
\left( |\varphi(\lam^2,\beta,\kappa)||\varphi(\lam_s^2,\beta,
\kappa)|^{- 1}\right) ^{\pm \eta(\lam_s-\lam)}\, , \]
and
\[ \delta_+(\lam)\delta_-(\lam)=(\lam-\lam_s)_+^{i\nu}(\lam-
\lam_s)_-^{i\nu} (\exp2i\gamma(\lam))\, , \]
showing integrability of the singularity at $\lam_s$.

\subsubsection{Estimation of $I^p(\lam)$ and $I^q(\lam)$}\label{estimsl}

\begin{figure}
\begin{center}
\includegraphics[scale=0.55]{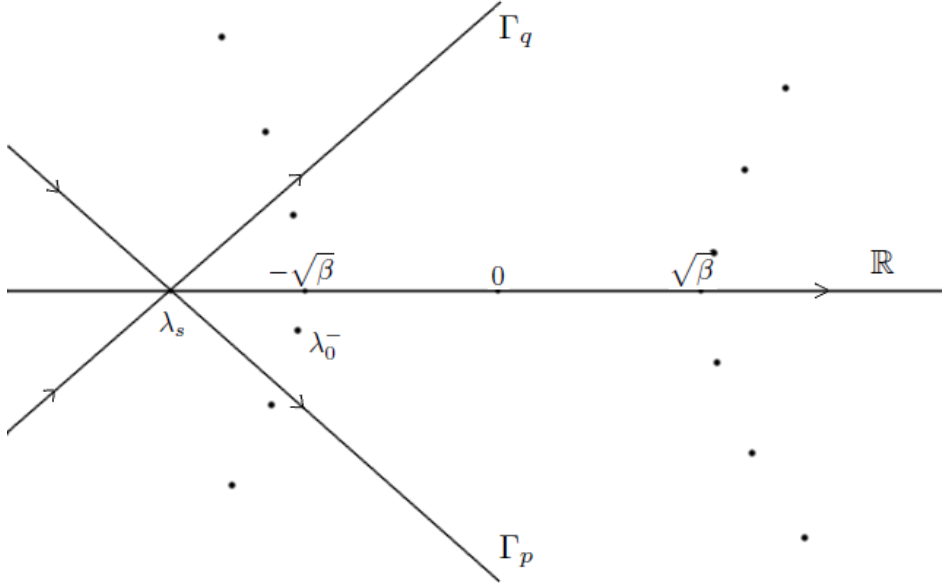}
\end{center}
\caption{Stationary-phase contours for evaluation of the \protect integrals
(\ref{aip}) and (\ref{aiq}) in the large-$t$ limit in the space-like case. The dots are the zeros of the function $e^{\lam^2-\beta}-e^{i\pi\kappa}$ with $\lam_0^-$ denoting the zero which gives the exponentially decreasing correction with the slowest rate of decay.} \label{figsl}
\end{figure}

In order to estimate $I^p(\lam)$ and $I^q(\lam)$ in the large-$t$
limit, we use the steepest-descent method to evaluate the integrals
(\ref{aip}) and (\ref{aiq}). The paths of the steepest descent going
through the stationary point $\lambda_s$ are shown in
Fig.~\ref{figsl}. An important consideration is that besides the
contribution to the integrals of this stationary point, which is of
the order
\be\label{m14} O\left(\frac{1}{\sqrt{t}(\lam-\lam_s)}\right) , \ee
one also has to take into account the contribution of the residues
located at $\lam\pm i0$ and at the zeros of the function
$e^{\lam^2-\beta}-e^{i\pi \kappa}$. We begin by first neglecting the
contributions from the residues at the zeros of
$e^{\lam^2-\beta}-e^{i\pi\kappa}$ which at large $t$ give
exponentially small corrections and focus on the residue at $\lam
\pm i0$. (A more complete estimate will be presented in the following
sections.) Transforming the integration contour in (\ref{aip}) from
the real axis to the steepest-descent path $\Gamma_p$ (see
Fig.~\ref{figsl}), and using the analytical properties of the
integrands discussed above, we obtain:
\be \label{Ipp}
I^p_+(\lam)=\eta(\lam_s-\lam)\delta_+(\lam)\delta_-(\lam)p(\lam)
e^{-2i\thet(x,t,\lam)}+O\left(\frac{1}{\sqrt{t}(\lam-\lam_s)}\right)
. \ee
Similarly,
\be\label{Ipm}
I^p_-(\lam)=-\eta(\lam-\lam_s)\delta_+(\lam)\delta_-(\lam)p(\lam)
e^{-2i\thet(x,t,\lam)}+O\left(\frac{1}{\sqrt{t}(\lam-\lam_s)}\right)
. \ee
For $I^q(\lam)$, the computations follow the same steps with the
steepest-descent path $\Gamma_q$ (see Fig.~\ref{figsl}), and the
result is:
\be\label{Iqpm} I^q_\pm(\lam)=\pm\eta(\mp\lam_s\pm \lam)
\delta_+^{-1} (\lam)\delta_-^{-1} (\lam) q(\lam)
e^{2i\thet(x,t,\lam)}+O\left(\frac{1}{\sqrt{t}(\lam-\lam_s)}
\right)\, . \ee

The calculations above are valid when $\lam$ is not too close to the
stationary point $\lam_s$. In the vicinity of the stationary point,
the integrals $I^p(\lam)$ and $I^q(\lam)$ can be estimated as
\[ I^p(\lam) \sim \frac{p_s}{2\pi i}\inti \frac{(\mu -\lam_s)_+^{i\nu}
(\mu -\lam_s)_-^{i\nu}\exp(-2i\thet(\mu))}{\mu-\lam}d\mu\, ,\]
and
\[ I^q(\lam) \sim \frac{q_s}{2\pi i}\inti \frac{(\mu -\lam_s)_+^{-i\nu}
(\mu -\lam_s)_-^{-i\nu}\exp(2i\thet(\mu))}{\mu-\lam}d\mu\, , \]
with
\[ p_s=p(\lam_s)\exp(2i\gamma(\lam_s))\, ,\ \ \ q_s=q(\lam_s)
\exp(-2i \gamma( \lam_s))\, . \]
This means that in the vicinity of $\lam_s$, the $t$-dependence of
$I^p$ and $I^q$ in the leading order is given by the following
relations:
\be\label{m15} I_s^p(\lam)= \inti \frac{|\mu|^{2 i \nu}
\exp(-2it\mu^2) }{\mu-(\lam - \lam_s )}d\mu= t^{-i\nu} \inti
\frac{|\mu|^{2 i \nu}\exp(-2i \mu^2)}{\mu-\sqrt{t}( \lam- \lam_s)}
d\mu\, , \ee
and
\be\label{m16} I_s^q(\lam)= \inti \frac{|\mu|^{-2 i \nu}
\exp(2it\mu^2)}{\mu- (\lam-\lam_s )}d\mu= t^{+i\nu} \inti
\frac{|\mu|^{-2 i \nu}\exp(2i\mu^2) }{\mu-\sqrt{t}( \lam- \lam_s)}
d\mu\, . \ee
The boundary values of these Cauchy integrals are uniformly bounded
(see \cite{G}) in $\sqrt{t}(\lam-\lam_s)$ due to the fact that $\nu$
is real. This proves that the boundary values of $I^p$ and $I^q$ and
therefore $\Phi^m_\pm(\lam)$ are bounded in the large-$t$ limit.

\subsection{Approximate solution of the RHP}

Now we are ready to show that the Manakov ansatz $\Phi^m(\lam)$ is
an approximate solution of the RHP (\ref{RHPT2}). More precisely, if
$\Phi(\lam)$ is the exact solution of (\ref{RHPT2}), then
\be\label{m19} \Phi(\lam)=[I+O(t^{-\varrho})]\Phi^m(\lam)\, ,\ \
\varrho \in\left(0,\frac{1}{2}\right) , \ \ \ \mbox{for}\;\; t \rightarrow +\infty\, ,\  -\frac{x}{2t}<-\sqrt{\beta}\, ,\; \beta>0 \, . \ee
Indeed, not too close to $\lam_s$, we have from Eqs.~(\ref{Ipp}),
(\ref{Ipm}), and (\ref{Iqpm}):
\[ [\Phi^m_+(\lam)]^{-1}=e^{-\sigma_3\ln\delta_+(\lam)}\left(
\begin{array}{cc} 1&\eta(\lam_s-\lam)\delta_+(\lam)\delta_-(\lam)
p(\lam)e^{-2i\thet(\lam)}\\ \eta(\lam-\lam_s) \delta_+^{-1}(\lam)
\delta_-^{-1}(\lam)q(\lam)e^{2i\thet(\lam)}&1
\end{array}\right)+O\left(\frac{1}{\sqrt{t}(\lam-\lam_s)}\right) ,
\]
and
\[ \Phi^m_-(\lam)=\left(\begin{array}{cc} 1& \eta(\lam-\lam_s)
\delta_+(\lam)\lam_-(\lam)p(\lam)e^{-2i\thet(\lam)}\\
\eta(\lam_s-\lam)\delta_+^{-1}(\lam)\delta_-^{-1}(\lam)q(\lam)
e^{2i\thet(\lam)}&1\end{array}\right) e^{\sigma_3\ln\delta_-(\lam)}
+O\left(\frac{1}{\sqrt{t}(\lam-\lam_s)}\right) , \]
so that
\[ [\Phi^m_+(\lam)]^{-1}\Phi^m_-(\lam)=\left(\begin{array}{cc}
1&p(\lam) e^{-2it\lam^2-2ix\lam}\\q(\lam) e^{2it\lam^2+2ix\lam}&
1+p(\lam)q(\lam)\end{array}\right)+
O\left(\frac{1}{\sqrt{t}(\lam-\lam_s)}\right). \]
Since the boundary values of (\ref{m15}) and (\ref{m16}) are
uniformly bounded, this means that
\be \label{m17}
[\Phi^m_+(\lam)]^{-1}\Phi^m_-(\lam)=G_\Phi(\lam)+r_0(\lam)\, , \ee
where $G_\Phi(\lam)$ is given by (\ref{m12}) and
\be\label{m18} r_0(\lam)=\left\{\begin{array}{l}
O\left(1/[\sqrt{t}(\lam-\lam_s)] \right)\, ,\ \ |\lam-\lam_s|>
t^{-1/2+\varrho}\, , \\ O(1)\, , \ \ |\lam-\lam_s|<
t^{-1/2+\varrho}\, , \end{array}\right. \ee
for $t\rightarrow\infty$ and with $\varrho \in (0,\frac{1}{2})$. If
one introduces the matrix
\[ R(\lam)\equiv \Phi(\lam)[\Phi^m(\lam)]^{-1}, \]
then
\[ R_+(\lam)-R_-(\lam)=\Phi_+(\lam)[\Phi^m_+(\lam)]^{-1}-
\Phi_-(\lam)[\Phi^m_-(\lam)]^{-1}, \]
and using Eq.~(\ref{m17}) and the relation $\Phi_+(\lam)
G_\Phi(\lam)= \Phi_-(\lam)$ we obtain
\[ R_+(\lam)-R_-(\lam)=r(\lam)\equiv\Phi_+(\lam)r_0(\lam)
[\Phi^m_-(\lam)]^{-1}\, . \]
Taking into account that $R(\infty)=I$ we see that this this
relation implies that the the matrix $R(\lam)$ can be represented
like this
\[ R(\lam)=I+\frac{1}{2\pi i}\inti\frac{r(\mu)}{\mu-\lam}d\mu\, ,
\ \ \ \mbox{for}\;\;  \lam\in\mathbb{C}/\mathbb{R}\, . \]
Under the hypothesis that $\Phi_+(\lam)$ is uniformly bounded in
$\lam\in\mathbb{R}$ (which can be proved as in \cite{IIKV}),
$r(\lam)$ satisfies the same estimates as $r_0(\lam).$ Therefore,
outside of a vicinity of $\lam_s$,
\[ R(\lam)=I+O(t^{-\varrho})\, ,\ \ \varrho\in(0,\frac{1}{2})\, , \]
proving Eq.~(\ref{m19}).

\subsection{Asymptotic behavior of the potentials}
\label{abp}

Making use of the Manakov ansatz
\[ \Phi^m(\lam)=\left(\begin{array}{cc} \delta(\lam)&-I^p(\lam)
\delta^{-1}(\lam)\\ -I^q(\lam)\delta(\lam)&\delta^{-1}(\lam)
\end{array}\right)\, , \]
one can extract the auxiliary potentials from the large-$\lam$
expansion using the formulae obtained in Section
\ref{potentials}. We start with $b_{++}$ which enters directly the
expression (\ref{deffcpot}) for the field correlator. Since $\Phi^m(\lam)$
is an approximate solution of the RHP, Eq.~(\ref{potsl}) can be written as
\be \label{m20} b_{++}= (\Phi^m_1)_{12}+o(1)=\frac{1}{2\pi
i}\inti\delta_+(\mu)\delta_-(\mu)p(\mu)e^{-2i\thet(x,t,\mu)}d\mu+o(1)\,
, \ee
with $p(\lam)$ given by Eq.~(\ref{defpspace}). The integral
appearing in this expression can be estimated via the
steepest-descent method in the same way as we did for $I^p(\lam)$ in
Section \ref{estimsl}. This means that if one neglects the
exponentially small corrections that come from the residues at the
zeros of $e^{\lam^2-\beta}-e^{i\pi\kappa}$, Eq.~(\ref{m20}) gives
\[ b_{++}= c_0 t^{-1/2-i\nu}e^{2it\lam_s^2}+o(1)\, , \]
where $\lam_s=-x/2t$, $\nu$ is defined by Eq.~(\ref{defnu}), and
$c_0$ is a constant which depends on $\beta$ and $\kappa$. Until
now, all the considerations were rigorous. The fact that the Manakov
ansatz is only an approximate solution of our RHP, as specified by
Eq.~(\ref{m19}), means then that the next term in the asymptotic
expansion could be of the order of $O(t^{-1/2-\varrho})$, and one
should not take into account the exponentially small terms which
appear in the complete evaluation of the integral (\ref{m20}). There
is, however, a caveat. The condition (\ref{c100}) ensures that the transformation of the integration contour from the real axis to the steepest-descent path encloses the pole at $\lam =\lam_0^-$ which is
closest to the real axis (see Fig.~\ref{figsl}) and is given by:
\[ \lam_0^-=-\left(\beta+\sqrt{\beta^2+\pi^2\kappa^2}\right)^{1/2}/
\sqrt{2}-i\left(-\beta+\sqrt{\beta^2+\pi^2\kappa^2}\right)^{1/2}/
\sqrt{2}\, . \]
The residue at $\lam =\lam_0^-$ gives the most slowly-decaying  exponential term  to the integral (\ref{m20}), i.e., a more complete solution for (\ref{m20})
including the contribution from $\lam =\lam_0^-$ is
\bea\label{m21}
b_{++} &= & c_0 t^{-1/2-i\nu}e^{2it\lam_s^2}+c_1e^{-2i\thet(x,y,\lam_0^-)}+o(1)\, ,\nonumber\\ &= &  c_0 t^{-1/2-i\nu} e^{2it\lam_s^2}+ c_1 e^{2t(\pi\kappa -i\beta)}e^{-2ix\lam_0^-}+o(1)\, . \eea
As one approaches the bosonic limit $\kappa=0$, the second term in (\ref{m21}) which arises from the pole at $\lam =\lam_0^-$  becomes dominant  even compared with a possible  $O(t^{-\varrho})$ term, since $\Im \lam_0^- \rightarrow 0$ in this limit. This term is the main component of $b_{++}$ in the case of impenetrable bosons. This  shows that the exact solution for $b_{++}$ should be written as
\be\label{bppsl}
b_{++}=   c_0 t^{-1/2-i\nu}e^{2it\lam_s^2}+\cdots+ c_1e^{2t(\pi\kappa-i\beta)}e^{-2ix\lam_0^-}+\cdots\, .
\ee
where the dots between $c_0$ and $c_1$ mean that there might be terms of order $O(t^{-1/2-\varrho})$  which are, however, smaller than the $c_1$ term when $\kappa \rightarrow 0$.

Although we will not use it below, we present the result for the potential $B_{--}$ which is
\[ B_{--}=- (\Phi^m_1)_{21}+o(1)=-\frac{1}{2\pi i} \inti \delta_+^{-1}(\mu) \delta_-^{-1}(\mu)p(\mu)e^{2i\thet(x,t,\mu)}d\mu+o(1)\, . \]
This means that
\be\label{m26}
B_{--}= c_0 t^{-1/2+i\nu}e^{-2it\lam_s^2}+o(1)\, .
\ee
By contrast, the results for potentials $B_{+-}$ and $C_{+-}+C_{-+}+B_{--}G$ will be very important for the subsequent calculations. They are:
\[ B_{+-}=-\alpha_0-(\Phi^m_1)_{11}+o(1)=-\alpha_0-\delta_0+o(1)\, , \]
and
\[ C_{+-}+C_{-+}+B_{--}G=(\Phi^m_2)_{22}-(\Phi^m_2)_{11}+o(1)=-2\delta_1 +o(1) \, , \]
where $\alpha_0$ is defined  by Eq.~(\ref{defalpha0}), and
\[  \delta_0=\frac{i}{\pi}\int_{-\infty}^{\lam_s} d\mu\ln|\varphi(\mu^2, \beta,\kappa)|\, , \ \ \ \delta_1=\frac{i}{\pi}\int_{-\infty}^{\lam_s} d\mu\ \mu \ln|\varphi(\mu^2,\beta,\kappa)|\, . \]
Introducing the function $I(\beta,\kappa)$:
\[ I(\beta,\kappa)=\Im\left(\inti d\mu \ \ln \varphi(\mu^2,\beta,\kappa) \right) , \]
we can rewrite the result for $B_{+-}$ as
\be\label{m22}
B_{+-}=-\frac{I(\beta,\kappa)}{2\pi}+\frac{i}{2\pi}\inti\mbox{sign}(\mu- \lam_s)\ln|\varphi(\mu^2,\beta,\kappa)|\ d\mu+o(1)\, . \ee
Also, using the fact that $\varphi(\mu^2,\beta,\kappa)$ is even function of $\mu$, one can transform the result for $C_{+-}+C_{-+}+B_{--}G$ into
\be\label{m23}
C_{+-}+C_{-+}+B_{--}G=\frac{i}{\pi}\inti \mbox{sign}(\mu-\lam_s)\mu \ln |\varphi(\mu^2,\beta,\kappa)|\ d\mu+o(1)\, .
\ee
%

\subsection{Asymptotic behavior of $\sigma(x,t,\beta,\kappa)$}
\label{asig}

The formulae (\ref{m22}) and (\ref{m23}) allow us to obtain the asymptotic expression for $\sigma$. As a first step, combining them with the differential equations (\ref{desigma}) we have:
\be\label{m24}
\6_x\sigma=\frac{i}{\pi}{\cal I}(\beta,\kappa)+\frac{1}{\pi} \inti \mbox{sign}(\mu-\lam_s)\ln|\varphi(\mu^2,\beta,\kappa)|\ d\mu+o(1)\, ,
\ee
and
\be\label{m25}
\6_t\sigma=\frac{2}{\pi}\inti \mbox{sign}(\mu-\lam_s)\mu \ln |\varphi(\mu^2,\beta,\kappa)|\ d\mu+o(1)\, . \ee
The asymptotic expression for $\sigma$ is obtained  integrating Eqs.~(\ref{m24}) and (\ref{m25}) over $x$ and $t$. This implies, however, that more accurate expressions for the derivatives of $\sigma$ that include the higher-order asymptotic terms are needed to have the same accuracy for $\sigma$ as for $b_{++}$ (\ref{bppsl}). To obtain these expressions we first note that Eq.~(\ref{m22}) implies that
\be\label{m27}
\6_x B_{+-}= \frac{i}{2\pi} \ln|\varphi(\lam_s^2,\beta,\kappa)| \frac{1}{t} =O\left(\frac{1}{t}\right)\, .
\ee
Combined with the first part of Eq.~(\ref{bpmx}), $\6_xB_{+-}=2ib_{++} B_{--}$, this result agrees with the estimates $b_{++}=O(1/t^{1/2})$ and $B_{--}=O(1/t^{1/2})$ that were already obtained in the previous section, see Eqs.~(\ref{m21}) and (\ref{m26}). Also, we know that the potentials $B_{--}$ and $b_{++}$ solve the separated nonlinear Schr\"{o}dinger equation (\ref{NSSE}) for which the general structure of the decreasing solutions is
(see, e.g., \cite{AS,IIKV}):
\be\label{aebpp}
b_{++}=t^{-1/2}\left(u_0+\sum_{n=1}^\infty\sum_{k=0}^{2n}\frac{(\ln 4t)^k}{t^n}u_{nk}\right)e^{2it\lam_s^2-i\nu\ln 4t}\, , \ee
\be\label{aebmm}
B_{--}=t^{-1/2}\left(v_0+\sum_{n=1}^\infty\sum_{k=0}^{2n}\frac{(\ln 4t)^k}{t^n}v_{nk}\right)e^{-2it\lam_s^2+i\nu\ln 4t}\, , \ee
where $u_0,v_0,u_{nk},v_{nk}$ and $\nu$ are  functions of $\lam_s=-x/2t.$ The parameters $\nu,u_{nk},v_{nk}$ can be expressed in terms of $u_0,\, v_0$. In particular,
\bea
v_{12}u_0+u_{12}v_0&=&0\, , \ \ \ \nu=-4u_0v_0\, , \nonumber\\
v_{11}u_0+u_{11}v_0&=&\frac{(\nu^2)''}{32}\, ,\\
v_{10}u_0+u_{10}v_0&=&\frac{(\nu\nu')'}{16}+\frac{i}{8}(v_0'u_0-v_0u_0')'\, ,\nonumber \eea
where the prime denotes the derivative with respect to $\lam_s$.
Now we can improve the asymptotic expansions for the derivatives $\6_x \sigma$ and $\6_t \sigma$. Substitution of (\ref{aebpp}) and (\ref{aebmm}) into $\6_xB_{+-}=2ib_{++}B_{--}$ gives
\be\label{m28}
\6_x B_{+-}=-\frac{i\nu}{2t}+i\frac{(\nu^2)''}{16}\frac{\ln 4t}{t^2}+ i \left[\frac{(\nu^2)''}{16}+\frac{i}{4}(v_0''u_0-v_0u_0'')\right]\frac{1}{t^2}
+O\left(\frac{\ln^4 4t}{t^3}\right). \ee
Comparing the first term in this expansion with Eq.~(\ref{m27}), we see that $ \nu= -\frac{1}{\pi} \ln |\varphi (\lam_s^2,\beta,\kappa)|>0$ in Eqs.~(\ref{aebpp}) and (\ref{aebmm}), in agreement with our previous notation (\ref{defnu}). Integrating Eq.~(\ref{m28}) over $x$  and using the first equation in (\ref{desigma}), we obtain
\[ \6_x\sigma=\frac{i}{\pi}I(\beta,\kappa)+ \frac{1}{\pi} \inti \mbox{sign} (\mu-\lam_s)\ln|\varphi(\mu^2,\beta,\kappa)|\ d\mu \]
\be
-\frac{(\nu^2)'}{4}\frac{\ln 4t}{t}-\left[\frac{(\nu^2)'}{4}+ i (v_0'u_0- v_0u_0') \right]\frac{1}{t}+O\left(\frac{\ln^4 4t}{t^2}\right)\, . \ee
Equation (\ref{e37}) can be rewritten as
\[ \6_x(C_{+-}+C_{-+}+B_{--}G)=b_{++}\6_xB_{--}-B_{--}\6_xb_{++}\, . \]
Then the asymptotic expansions (\ref{aebpp}) and (\ref{aebmm}) give
\[ \6_x(C_{+-}+C_{-+}+B_{--}G)=i\frac{x\nu}{2t^2}+ \frac{ix\6_x((\nu^2)') }{8t^2} -\frac{x\6_x(v_0'u_0-u_0'v_0)}{2t^2}-\frac{(u_0v_0'-v_0u_0')}{2t^2} \]
\be\label{sx}
+i\frac{x\6_x((\nu^2)')}{8}\frac{\ln 4t}{t^2}-i\frac{\nu(\6_x\nu)}{2} \frac{\ln 4t}{t}+ O\left(\frac{\ln^4 4t}{t^3}\right)\, . \ee
Integrating this equation over $x$, and using the second equation in (\ref{desigma}) we find
\[  \6_t \sigma=\frac{2}{\pi}\inti \mbox{sign}(\mu-\lam_s)\mu \ln |\varphi(\mu^2,\beta,\kappa)|\ d\mu-\lam_s\frac{(\mu^2)'}{2}\frac{\ln 4t}{t}+ \frac{\nu^2}{2t} \]
\be\label{st}
-\lam_s\frac{(\nu^2)'}{2t}-\lam_s\frac{2i}{t}(v_0'u_0-v_0u_0') +O\left(\frac{\ln^4 4t}{t^2}\right)\, . \ee
Finally, integration of Eq.~(\ref{sx}) over $x$ and (\ref{st}) over $t$ gives the asymptotic expansion for $\sigma(x,t,\beta,\kappa)$ of required accuracy:
\[ \sigma(x,t,\beta,\kappa)=x\frac{i}{\pi}I(\beta,\kappa)+\frac{1}{\pi} \inti |x+2t|\ln|\varphi(\mu^2,\beta,\kappa)|\ d\mu+\frac{\nu^2}{2} +\frac{\nu^2}{2}\ln 4t \]
\be\label{sigmasl}
 +2i\int_{-\infty}^{\lam_s}(v_0'(\mu)u_0(\mu)-v_0(\mu)u_0'(\mu)\ d\mu+c(\beta)+O\left(\frac{\ln^4 4t}{t}\right)\, , \ee
where $c(\beta)$ is a constant that depends only on $\beta$.

\section{Asymptotic solution of the RHP. Time-like case}\label{RHPtime}

The computations in the time-like region defined by
\[ \lam_s>-\sqrt{\beta}\, ,\  \beta=h/T>0\, , \]
are very similar to those presented above for the space-like case. Because of this, the presentation in this Section is more sketchy, emphasizing the
differences between the two regions. As we will see in what follows,  the leading term of the asymptotics for the potential $b_{++}$ is the same in
the time-like as in the space-like region. The sub-leading term in the asymptotic expansion, which is important because  it reproduces the
 predictions of  conformal field theory, is, however, different in the time-like case.

\subsection{Manakov ansatz}

The Manakov ansatz in the time-like region is:
\be\label{MAT}
\Phi^m(\lam)=\left(\begin{array}{cc} 1&-I^p(\lam)\\
-I^q(\lam)&1 \end{array}\right)e^{-\sigma_3\ln\delta(\lam)} \, , \ee
where $I^p(\lam)$ and $I^q(\lam)$ are now given by
\be \label{ipt}  I^p(\lam)=\frac{1}{2\pi i}\inti \frac{\delta_+^{-1} (\mu)\delta_-^{-1} (\mu)}{ \mu-\lam}p(\mu)e^{-2i\thet(x,t,\mu)}d\mu\, , \ee
and
\be \label{iqt} I^q(\lam)=\frac{1}{2\pi i}\inti \frac{\delta_+(\mu)
\delta_- (\mu)}{\mu-\lam}q(\mu)e^{2i\thet(x,t,\mu)}d\mu\, . \ee
The functions $p(\lam)$ and $q(\lam)$ here are defined in Eqs.~(\ref{defptime}) and (\ref{defqtime}), respectively, while the function $\delta(\lam)$ is the solution of the following scalar RHP
\begin{equation*}
\delta_+(\lam)=\delta_-(\lam)[1+p(\lam)q(\lam)\eta(\lam-\lam_s)]\, ,\ \lam\in\mathbb{R}\, ,\ \ \ \ \delta(\infty)=1\, .
\end{equation*}
Using the fact that for $p(\lam)$ and $q(\lam)$ defined by (\ref{defptime}) and (\ref{defqtime}), $1+p(\lam)q(\lam)=|\varphi(\lam^2,\beta,-\kappa)|^2$, one can see that solution of this RHP can be written as
\[ \delta(\lam)=\exp\left\{\frac{1}{2\pi i}\int_{\lam_s}^{\infty} \frac{d\mu}{\mu-\lam}\ln|\varphi(\mu^2,\beta,-\kappa)|^2\right\}\, . \]

\subsubsection{Properties of\  $\delta(\lam)$}

Following the same steps as in the space-like region, we have
\be
\delta_{\pm}(\lam)=(\lam-\lam_s)_{\pm}^{-i\nu}\exp(i\gamma(\lam)) |\varphi(\lam_s^2,\beta,-\kappa)|^{\pm} \left(|\varphi (\lam^2,\beta, -\kappa)|\right)^{\pm \eta(\lam-\lam_s)}\, , \ee
with
\[\nu(\lam_s,\beta,\kappa)=- \frac{1}{\pi}\ln|\varphi(\lam_s^2, \beta, \kappa)| =- \frac{1}{\pi}\ln|\varphi(\lam_s^2,\beta,-\kappa)|>0\, , \]
and
\[ \gamma(\lam)=\frac{1}{\pi}\int_{\lam_s}^{\infty}\ln|\mu-\lam|d (\ln|\varphi(\mu^2,\beta,-\kappa)|)d\mu\, . \]
This also means that
\be
\delta_+(\lam)\delta_-(\lam)=(\lam-\lam_s)_+^{-i\nu}(\lam-\lam_s)_-^{-i\nu}  (\exp2i\gamma(\lam))\, , \ee
showing integrability of the singularity at $\lam_s$.

\subsubsection{Estimation of $I^p(\lam)$ and $I^q(\lam)$}

\begin{figure}
\begin{center}
\includegraphics[scale=0.55]{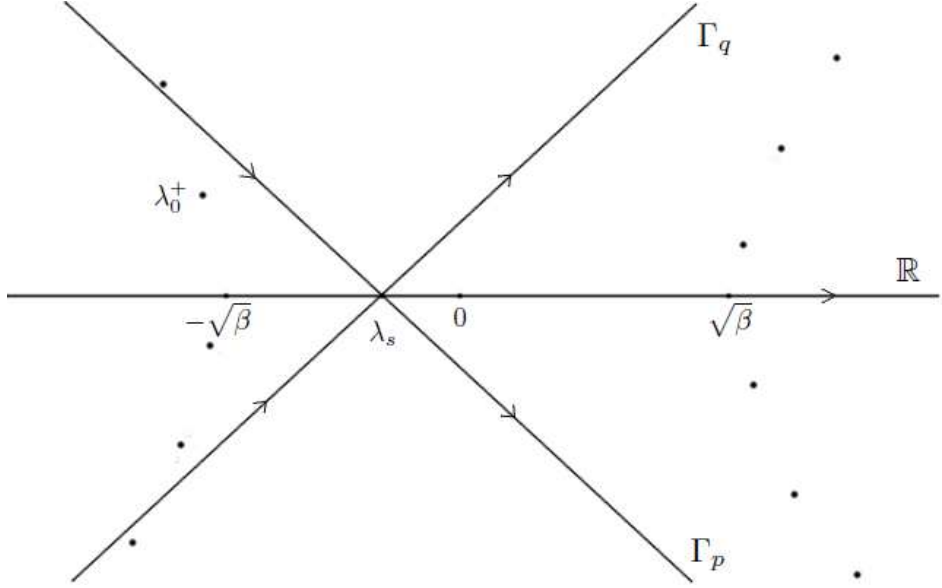}
\end{center}
\caption{Stationary-phase contours for the \protect integrals (\ref{ipt}) and (\ref{iqt}) in the large-$t$ limit in the time-like case. The dots are the zeros of the function $e^{\lam^2-\beta}-e^{-i\pi\kappa}$ with $\lam_0^+$ denoting the zero which gives the exponentially decreasing correction with the slowest rate of decay for $b_{++}$ in the time-like case.}\label{figtl}
\end{figure}

The estimates of $I^p(\lam)$ and $I^q(\lam)$ in the time-like region are obtained as in the space-like region by the steepest-decent method. The steepest-descent contours are shown in Fig.~\ref{figtl}.
Similarly to the space-like case, for $\lam$ not too close to the stationary point $\lam_s$, transformation of the integration contour from the real axis to the steepest-descent paths gives:
\be\label{ippmt}
I^p_\pm(\lam)=\pm \eta(\pm\lam_s\mp\lam)\delta_+^{-1}(\lam)\delta_-^{-1} (\lam) p(\lam)e^{-2i\thet(x,t,\lam)}+ O\left(\frac{1}{\sqrt{t}(\lam-\lam_s)} \right) ,\ee
and
\be\label{iqpmt}
I^q_\pm(\lam)=\pm\eta(\mp\lam_s\pm \lam)\delta_+(\lam)\delta_-(\lam) q(\lam) e^{2i\thet(x,t,\lam)}+O\left(\frac{1}{\sqrt{t}(\lam-\lam_s)}\right) .\ee
Also similarly to the space-like case, one can show that for $\lam$ in the vicinity of the stationary point, the boundary values of $I^p$ and $I^q$,  and therefore $\Phi^m_\pm(\lam)$, are bounded in the large-$t$ limit.

\subsection{Approximate solution of the RHP }

Combining Eqs.~(\ref{ippmt}), (\ref{iqpmt}), and (\ref{MAT}), one obtains
\[ [\Phi^m_+(\lam)]^{-1}=e^{\sigma_3\ln\delta_+(\lam)} \left( \begin{array}{cc} 1&\eta(\lam_s-\lam)\delta_+ (\lam)^{-1} \delta_-^{-1}(\lam) p(\lam)e^{-2i\thet(\lam)} \\
\eta(\lam-\lam_s)\delta_+(\lam)\delta_(\lam)q(\lam)e^{2i\thet(\lam)}&1
\end{array}\right)+O\left(\frac{1}{\sqrt{t}(\lam-\lam_s)}\right) , \]
and
\[ \Phi^m_-(\lam)=\left(\begin{array}{cc} 1& \eta(\lam-\lam_s) \delta_+(\lam)^{-1}\lam_-^{-1}(\lam)p(\lam)e^{-2i\thet(\lam)}\\
\eta(\lam_s-\lam)\delta_+(\lam)\delta_-(\lam)q(\lam)e^{2i\thet(\lam)}&1
\end{array}\right)e^{-\sigma_3\ln\delta_-(\lam)}+O\left(\frac{1}{\sqrt{t} (\lam-\lam_s)}\right) , \]
and therefore,
\[ [\Phi^m_+(\lam)]^{-1}\Phi^m_-(\lam)=\left(\begin{array}{cc} 1+p(\lam)q(\lam)&p(\lam) e^{-2it\lam^2-2ix\lam}\\ q(\lam) e^{2it\lam^2+ 2ix\lam}&1 \end{array}\right) +O\left(\frac{1}{\sqrt{t} (\lam- \lam_s)}\right) . \]
This shows that the  Manakov ansatz (\ref{MAT}) is an approximate solution for the RHP (\ref{RHPT2}) with the conjugation matrix (\ref{g1}).  More precisely, if $\Phi(\lam)$ is the exact solution, then
\[ \Phi(\lam)=[I+O(t^{-\varrho})]\Phi^m(\lam)\, ,\ \  \varrho\in\left(0,\frac{1}{2}\right) ,\ \ \ \mbox{for}\;\; t\rightarrow +\infty\, , \; -\frac{x}{2t}> -\sqrt{\beta}\, ,\; \beta>0\, . \]
%

\subsection{Asymptotic behavior of the potentials}

As above, the asymptotic expressions for the potentials are extracted from the large-$\lam$ expansion of Eq.~(\ref{MAT}) making  use of the formulae obtained in Section \ref{potentials}. Substituting Eqs.~(\ref{MAT}) and (\ref{ipt}) into the second equation in (\ref{potsl}), we have
\[ b_{++}= (\Phi^m_1)_{12}+o(1)=\frac{1}{2\pi i}\inti\delta_+(\mu)^{-1} \delta_-(\mu)^{-1}p(\mu)e^{-2i\thet(x,t,\mu)}d\mu+o(1)\, , \]
where $p(\lam)$ is given by Eq.~(\ref{defptime}). The main difference with the space-like region is that now the residues that give the exponential corrections to the leading term of the asymptotics are the zeros of the function $e^{\lam^2-\beta}-e^{-i\pi\kappa}$ and not $e^{\lam^2-\beta}- e^{i\pi\kappa}$. The pole that is closest to the real axis among those that are enclosed by the transformation of the integration contour from the real axis to $\Gamma_p$ (see Fig.~\ref{figtl}), and therefore contributes the most slowly-decaying exponential term, is:
\[ \lam_0^+=-\left(\beta+\sqrt{\beta^2+\pi^2\kappa^2}\right)^{1/2}/\sqrt{2} +i\left(-\beta+\sqrt{\beta^2+\pi^2\kappa^2}\right)^{1/2}/\sqrt{2}\, . \]
The contributions of the stationary point and the residue at $\lam_0^+$ produce then the following expression for $b_{++}$:
\bea\label{bpptl}
b_{++} &= & c_0 t^{-1/2-i\nu}e^{2it\lam_s^2}+c_1e^{-2i\thet(x,y,\lam_0^-)}+o(1)\, ,\nonumber\\ &= &  c_0 t^{-1/2-i\nu} e^{2it\lam_s^2} +c_1 e^{2t(-\pi\kappa-i\beta)}e^{-2ix\lam_0^-}+o(1)\, , \eea
where $c_0$ and $c_1$ are some undetermined amplitudes which can depend on $\beta$, $\kappa$, and $\lam_s$. Again, as we approach the bosonic limit, $\kappa \rightarrow 0$, the second term in (\ref{bpptl}) becomes dominant. This term represents the leading asymtoptic term of $b_{++}$ for impenetrable bosons.

The leading term of the potential $B_{--}$ is given by the same Eq.~(\ref{m26}) as in the space-like case. The potentials $B_{+-}$ and $C_{+-}+C_{-+}+B_{--}G$ are obtained from Eq.~(\ref{pottl})
\[ B_{+-}=+\alpha_0-(\Phi^m_1)_{11}+o(1)=+\alpha_0-\delta_0+o(1)\, , \]
and
\[ C_{+-}+C_{-+}+B_{--}G=(\Phi^m_2)_{22}-(\Phi^m_2)_{11}+o(1) =-2\delta_1+o(1)\, , \]
where $\alpha_0$ is now defined by Eq.~(\ref{defalpha0t}), and
\be
\delta_0=\frac{1}{\pi i}\int_{\lam_s}^{\infty} d\mu \ln| \varphi(\mu^2, \beta,-\kappa)|\, , \ \ \ \delta_1=\frac{1}{\pi i}\int_{\lam_s}^{\infty} d\mu\ \mu \ln|\varphi(\mu^2,\beta,-\kappa)|\, . \ee

The result for $B_{+-}$ can be rewritten as
\be\label{m29}
B_{+-}=\frac{I(\beta,-\kappa)}{2\pi}+\frac{i}{2\pi}\inti\mbox{sign} (\mu-\lam_s)\ln|\varphi(\mu^2,\beta,-\kappa)|\ d\mu+o(1)\, . \ee
in terms of the function
\[ I(\beta,-\kappa)=\Im\left(\inti d\mu \ \ln \varphi(\mu^2, \beta, -\kappa)\right)=-\Im\left(\inti d\mu \ \ln \varphi(\mu^2,\beta, \kappa)\right). \]
Also, using the fact that $\varphi(\mu^2,\beta,-\kappa)$ is an even function of $\mu$, the equation for $C_{+-}+C_{-+}+B_{--}G$ can be transformed into
\be\label{m31}
C_{+-}+C_{-+}+B_{--}G=\frac{i}{\pi}\inti \mbox{sign}(\mu-\lam_s)\mu \ln |\varphi(\mu^2,\beta,-\kappa)|\ d\mu+o(1)\, . \ee
%

\subsection{Asymptotic behavior of $\sigma(x,t,\beta,\kappa)$}

Taking into account that $I(\beta,-\kappa)=-I(\beta,\kappa)$, and $|\varphi(\mu^2,\beta,-\kappa)|=|\varphi(\mu^2,\beta,\kappa)|$, one sees directly that the
 asymptotic expressions (\ref{m29}) for $B_{+-}$ and (\ref{m31}) for $C_{+-}+C_{-+}+B_{--}G$ in the time-like case coincide with the corresponding expressions (\ref{m22}) and (\ref{m23}) in the space-like region. Since the higher-order corrections discussed in Sec.~\ref{asig} are the same in both regions, this means that the asymptotic expansion for $\sigma$ in the time-like case is given by the same Eq.~(\ref{sigmasl}) as before.

\section{Results}\label{results}

Now we have all the ingredients to formulate the results for the main object of our interest, the anyonic field correlator which, as a reminder, is given in rescaled variables by the expression
\be \la\fa(x_2,t_2)\fad(x_1,t_1)\ra_T=\sqrt{T}g(x,t,\beta,\kappa) \, \ee
with
\be
g(x,t,\beta,\kappa)=-\frac{1}{2\pi}e^{2it\beta}b_{++}(x,t,\beta,\kappa) e^{\sigma(x,t,\beta,\kappa)}\, . \ee
%

\subsection{Negative Chemical Potential}

While all the considerations in the previous sections were based on the assumption that the chemical potential is positive, in fact, the results obtained are also valid when $\beta<0$. In this case, we need only the leading term for $b_{++}$. Putting together the first term in Eq.~(\ref{bppsl}) or (\ref{bpptl}) and Eq.~(\ref{sigmasl}), we can express the leading asymptotic behavior of the anyonic field correlator at negative chemical potential as
\be\label{asymptncp}
g(x,t,\beta,\kappa)= c_0 t^{(\nu-i)^2/2}e^{2it(\lam_s^2+\beta)}e^{ix I(\beta,\kappa)/\pi}e^{C(x,t,\beta,\kappa)}\left[1+ o\left(t^{-1/2} \right)\right] , \ee
where $c_0$ is some constant amplitude, $\lam_s=-x/2t$, $\nu=-(1/\pi)\ln|\varphi(\lam_s^2,\beta,\kappa)|$, and the definitions of all other functions in this equation are presented together in Eqs.~(\ref{int2}) and (\ref{int3}) of the Introduction.

\subsection{Positive Chemical Potential}

For reasons discussed in Sec.~\ref{abp}, in the case of positive chemical potential, one needs to keep in the asymptotic expansion of the potential    $b_{++}$ and, therefore, of the field correlator, not only the leading term, which is the same in the space-like and time-like regions, but the next exponentially decreasing term as well, which is different in the two regions. Thus, the two results should be presented separately.

{\bf Space-Like Region: $x/2t>\sqrt{\beta}$.} Combining Eqs.~(\ref{sigmasl}) and (\ref{bppsl}), we have
\be\label{asymppcps}
g(x,t,\beta,\kappa)= t^{\nu^2/2}e^{ix I(\beta,\kappa)/\pi} e^{C(x,t,\beta,\kappa)} \left[c_0 t^{-1/2-i\nu}e^{2it(\lam_s^2+\beta)}+
c_1e^{2t\pi\kappa}e^{-2ix\lam_0^-}+o\left(t^{-1/2}\right)\right] .
\ee

{\bf Time-Like Region: $x/2t<\sqrt{\beta}$.} In this case, Eqs.~(\ref{sigmasl}) and (\ref{bpptl}) give:
\be\label{asymppcpt}
g(x,t,\beta,\kappa)=  t^{\nu^2/2} e^{ix I(\beta,\kappa)/\pi} e^{C(x,t,\beta,\kappa)}\left[c_0 t^{-1/2-i\nu}e^{2it(\lam_s^2+\beta)}+
c_1e^{-2t\pi\kappa}e^{-2ix\lam_0^+}+o\left(t^{-1/2}\right)\right] .
\ee
The constants $\lam_0^-$ and $\lam_0^+$ in these equations are defined by
Eq.~(\ref{int5}) of the Introduction.

\section{Bosonic and Free-Fermionic Limit}\label{bflimit}
As the last step, we analyze our main result for the anyonic field correlator in various limits, in order to establish the relation with previously known
expressions, and to demonstrate the unexpected features of the anyonic case.

{\bf Bosonic Limit. } For bosons, $\kappa\rightarrow 0$, one has:
\be
\varphi(\lam^2,\beta,\kappa=0)=\frac{e^{\lam^2-\beta}-1}{e^{\lam^2-\beta}+1} \, ,\ \ C(x,t,\beta,\kappa=0)=\frac{1}{\pi}\inti|x-2t\lam|\ln |\varphi( \lam^2,\beta,\kappa=0)|\ d\lam\, , \ee
and $\nu=-(1/\pi)\ln|\varphi(\lam_s^2,\beta,\kappa=0)|$. Also, $I(\beta,\kappa=0)=-2\pi\sqrt{\beta}$ for $\beta>0$, and $I(\beta,\kappa=0)=0$ for $\beta<0$. In the case of negative chemical potential, using these relations it is straightforward to see that Eq.~(\ref{asymptncp}) reduces to the known result for impenetrable bosons \cite{IIKV,KBI}. For positive chemical potential, the result obtained in \cite{IIKV,KBI} is
\be
g(x,t,\beta,\kappa=0)=c_0t^{\nu^2/2}e^{C(x,t,\beta,\kappa=0)}\left[1+ O(t^{-1/2})\right] ,\ee
and is valid in both the space-like and the time-like region. Taking into account that in both regions, $\lam_0^\pm=-\sqrt{\beta}$ for $\kappa=0$,
we can see that in Eqs.~(\ref{asymppcps}) and (\ref{asymppcpt}), the second term in the parenthesis gives the leading contribution in this limit, which reproduces the bosonic result. This means that for a certain value of $\kappa$ approaching 0, there is a crossover in which the relative magnitude of the two terms in the parenthesis changes, and the second term becomes
the leading one for $\kappa$ close to 0.

{\bf Free-Fermionic Limit.} For $\kappa\rightarrow 1$, the anyonic system we considered reduces to free fermions. In this case,
the function $\varphi(\lam^2,\beta,\kappa)$ vanishes, which means that $\nu=0$ and $C(x,t,\beta,\kappa=1)=0$. It is easy to see that
 Eqs.~(\ref{asymppcps}) and (\ref{asymppcpt}) reduce to the corresponding correlators (\ref{asympffs}) and (\ref{asympfft}) of free
 fermions that are presented in Appendix \ref{ferm}.

\section{Conformal Field Theory}\label{CFT}

The behavior of the field-field correlators of the one-dimensional particle systems at low temperatures is usually believed to follow the predictions of  conformal field theory (CFT). For impenetrable anyons, the CFT result for the leading term of the large time and distance asymptotic of the field-field correlator is \cite{PKA}:
\[ \la \fa(x,t)\fad(0,0)\ra \sim e^{-ik_F\kappa x}\exp\left\{-\left(\frac{2\pi T \Delta^+}{v_F}|x-v_Ft|+\frac{2\pi T \Delta^-}{v_F}|x+v_Ft|\right)\right\}\, , \]
where  $k_F=\sqrt{h}$ and $v_F=2k_F$ are the Fermi vector and the Fermi velocity, respectively, and $\Delta^\pm$ are the conformal dimensions:
\[ 2\Delta^\pm=\left(\frac{1}{2{\cal Z}}\mp{\cal Z}\frac{\kappa}{2}\right)^2 ,\ \ \ {\cal Z}=1\, . \]
(Note the differences in conventions between \cite{PKA} and this work together with the rest of the papers \cite{PKA2,PKA3}, where
we have obtained the determinant representation for the anyonic field-field correlator. The main difference is related to the ordering of the anyonic creation operators in the eigenstates of the Hamiltonian, which amounts with the change of the sign of the statistics parameter $\kappa$ in the formulae of \cite{PKA}.) In the notations of this work, the field correlator  considered here is
\[ \la\fa(x_2,t_2)\fad(x_1,t_1)\ra_T\, , \;\;\; t_{21}=t_2-t_1>0\, , \;\;  x_{12}=x_1-x_2>0\, , \]
which means that the CFT predictions are
\be
\la\fa(x_2,t_2)\fad(x_1,t_1)\ra_T\sim e^{ik_F\kappa x_{12}}e^{\pi T\kappa t_{21}} e^{-\frac{\pi T } {v_F}x_{12}\left(\frac{1}{2} +\frac{\kappa^2}{2} \right)}, \ee
in the space-like region, and

\be
\la\fa(x_2,t_2)\fad(x_1,t_1)\ra_T\sim e^{ik_F\kappa x_{12}}e^{\frac{\pi T\kappa}{v_F}x_{12}} e^{-\pi T t_{21}\left(\frac{1}{2}+\frac{\kappa^2}{2} \right)}, \ee
in the time-like region.
The leading term of the asymptotics (\ref{asymppcps}) and (\ref{asymppcpt}) obtained from the exact calculation in this work {\em do not} reproduce these equations in the limit of low temperatures  $\beta\rightarrow \infty$. We can show, however, that the sub-leading terms in these asymptotics do
give the conformal behavior. Indeed, as shown in the Appendix \ref{analysisc}, in the limit $\beta\rightarrow \infty$, we have:
\be
x\frac{i}{\pi}I(\beta,\kappa)+C(x,t,\beta,\kappa)=x2i\sqrt{\beta}(\kappa-1) -x\frac{\pi}{2\sqrt{\beta}}(1-\kappa)^2,
\ee
in the space-like case, and
\be\label{m36}
x\frac{i}{\pi}I(\beta,\kappa)+C(x,t,\beta,\kappa)=x2i\sqrt{\beta}(\kappa-1) -t\pi (1-\kappa)^2 ,
\ee
in the time-like case. In the same limit,  $\lam_0^-$ and $\lam_0^+$ given by (\ref{int5}) become
\be\label{m37}
\lam_0^{\pm}=-\sqrt{\beta}\pm i\frac{\pi\kappa}{2\sqrt{\beta}}\, .
\ee
Using these formulae, we see directly that the second term in the asymptotic expansion of the field correlator is given by
\be
t^{\nu^2/2}e^{2ix\kappa\sqrt{\beta}}e^{2t\pi\kappa}e^{-x\frac{\pi}{2\sqrt{\beta}} (\kappa^2+1)},
\ee
in the space-like, and
\be
t^{\nu^2/2}e^{2ix\kappa\sqrt{\beta}}e^{x\frac{\pi\kappa}{\sqrt{\beta}}}e^{-t \pi (\kappa^2+1)},
\ee
in the time-like region. The exponential terms are exactly the ones predicted by CFT, if we take into account that $x=x_{12}\sqrt{T}$, $t=t_{21}T/2$, and $\beta=h/T$.

Qualitatively, the non-conformal term of the time-dependent field-field correlator, which is the leading asymptotic term for particle statistics not too close to bosons, can be traced back \cite{PKA6} to the singularity of the one-dimensional density of states at the bottom of the single-particle energy spectrum $\lambda \rightarrow 0$. In agreement with this interpretation, there is no non-conformal terms in the ``static'' equal-time correlator (see, e.g., \cite{PKA4}), since the single-particle spectrum is unlimited in momentum space, $\lambda \in (-\infty , +\infty)$. By contrast, the energy spectrum $\epsilon \propto \lambda^2$ has a threshold at $\lambda = 0$ with associated non-analytical behavior of the density of states. This non-analyticity manifests itself directly through the non-conformal terms in the asymptotic behavior of the field correlator of the massive one-dimensional particles.

\section{Summary}

In conclusion, we have calculated the large time and distance
asymptotic behavior of the temperature dependent field-field
correlation functions of impenetrable one-dimensional spinless
anyons. As a function of the statistics parameter, the anyonic
correlator interpolates continuously between the two limits of
impenetrable bosons and free fermions. The main qualitative feature
of our result is that, asymptotically, the anyonic correlator
consists of two additive parts. One is a non-conformal term produced
by the non-analyticity of the density of states at the bottom of the
single-particle energy spectrum. For all values of the particle
statistics away from the bosonic limit, this term gives the leading
asymptotic contribution to the correlator. The other is the
sub-leading term which agrees with the conformal field theory and is
associated physically with the low-energy excitations close to the
effective Fermi energy of the system of impenetrable anyons. In
agreement with the previous results \cite{KBI,EKL}, for the
statistics parameter close to the bosonic limit, the conformal term
determines the leading behavior of the asymptotics. Because of the
additivity of the two parts of the correlator and their different
physical origin, even away from the boson limit, the systems
response to the low-energy probes is determined by the (sub-leading)
conformal part of the correlator.

\appendix

\section{Large time and distance asymptotic behavior of the field
               correlator  for free fermions} \label{ferm}

In rescaled variables used in this work, $t=(t_2-t_1)T/2>0,\  x=(x_1-x_2)\sqrt{T}/2>0,\  \beta=h/T$, the field-field correlation function  of free one-dimensional fermions is expressed as \cite{AGD}
\[ \la \fa(x_2,t_2)\fad(x_1,t_1)\ra_T=\sqrt{T}\frac{e^{2it\beta}}{2\pi} \inti d\lam \frac{e^{\lam^2-\beta}}{e^{\lam^2-\beta}+1} e^{-2i\thet(x,t, \lam)} . \]
We are interested in the asymptotic behavior of the correlator in the limit of large  $x>0,\  t>0$ with $x/t=const$. The analysis is similar to the one performed for the functions $I^p(\lam)$ in Section \ref{estimsl}. The leading term is obtained via the steepest descent method and the corrections come from the poles located in the complex plane at the zeroes of the function $e^{\lam^2-\beta}+1$. The corrections to the leading term
are different in the space-like and time-like regions.

{\bf Space-like region: ($x/2t>\sqrt{\beta}$)}. In this case, the residue that gives the exponential term with the slowest rate of decay for large $x$ and $t$ is
\[ \lam_0^s=-\left(\beta+\sqrt{\beta^2+\pi^2}\right)^{1/2}/\sqrt{2}- i\left(-\beta+\sqrt{\beta^2+\pi^2}\right)^{1/2}/\sqrt{2}\, , \]
resulting in the following asymptotic behavior of the correlator:
\be\label{asympffs}
\la \fa(x_2,t_2)\fad(x_1,t_1)\ra_T\sim c_0t^{-1/2}e^{2it(\beta+\lam_s^2) }+c_1e^{2t\pi}e^{-2ix\lam_0^s}+\cdots\, .
\ee
Here $\lam_s=-x/2t$ is the stationary point of the phase $\thet(x,t,\lam)$ and $c_0, c_1$ are some constant amplitudes. In the limit of low temperatures, $\beta\rightarrow \infty$, using the fact that $\lam_0^s\rightarrow -\sqrt{\beta}-i\pi/(2\sqrt{\beta})$, we obtain
\[ \la \fa(x_2,t_2)\fad(x_1,t_1)\ra_T\sim c_0t^{-1/2}e^{2it(\beta +\lam_s^2) }+c_1e^{2ix \sqrt{\beta}} e^{2\pi(t-x/2\sqrt{\beta})}+\cdots\, . \]

{\bf Time-like region: ($x/2t<\sqrt{\beta}$)}. In this case, the residue producing the leading contribution is:
\[ \lam_0^t=-\left(\beta+\sqrt{\beta^2+\pi^2}\right)^{1/2}/\sqrt{2}+ i\left(-\beta+\sqrt{\beta^2+\pi^2}\right)^{1/2}/\sqrt{2}\, . \]
with the corresponding asymptotic behavior of the correlator:
\be\label{asympfft}
\la \fa(x_2,t_2)\fad(x_1,t_1)\ra_T\sim c_0t^{-1/2}e^{2it(\beta+\lam_s^2) }+c_1e^{-2t\pi}e^{-2ix\lam_0^t}+\cdots\, .
\ee
In the low-temperature limit, this becomes
\[ \la \fa(x_2,t_2)\fad(x_1,t_1)\ra_T\sim c_0t^{-1/2}e^{2it(\beta+ \lam_s^2) } +c_1e^{2ix \sqrt{\beta}} e^{2\pi(-t+x/2\sqrt{\beta})}+\cdots\, . \]
%

\section{Analysis of $C(x,t,\beta,\kappa)$}\label{analysisc}

The function $C(x,t,\beta,\kappa)$ is defined in the main text as
\be
C(x,t,\beta,\kappa)=\frac{1}{\pi}\inti|x-2t\lam|\ln|\varphi(\lam^2,\beta,\kappa)|\ d\lam\, , \ee
where
\be
\varphi(\lam^2,\beta,\kappa)=\frac{e^{\lam^2-\beta}-e^{i\pi\kappa}}{ e^{\lam^2-\beta}+1}\, . \ee
Using the expansion of the logarithm: $\ln(1-z)=-\sum_{n=1}^\infty z^n/n\, ,\  |z|<1$, we obtain the following expansions for $\ln|\varphi (\lam^2, \beta,\kappa)|$:
\be\label{m311}
\ln|\varphi(\lam^2,\beta,\kappa)|=-\sum_{n=1}^\infty \frac{e^{n(\lam^2-\beta)}}{n}\left(\cos(n\pi\kappa)+(-1)^{n+1}\right), \;\;\; \lam\in(-\sqrt{\beta},\sqrt{\beta})\, ,
\ee
and
\be\label{m32}
\ln|\varphi(\lam^2,\beta,\kappa)|=-\sum_{n=1}^\infty \frac{e^{n(\beta-\lam^2)}}{n}\left(\cos(n\pi\kappa)+(-1)^{n+1}\right) ,
\;\;\; \lam\in (-\infty,-\sqrt{\beta})\cup(\sqrt{\beta},\infty) \, .  \ee
We are interested in the the asymptotic behavior of $C(x,t,\beta,\kappa)$ in the limit of low temperatures $(\beta\rightarrow \infty)$. This behavior is different in the space-like and time-like region.

{\bf Space-like region: ($x/2t>\sqrt{\beta}$)}. It is convenient to express the function $C(x,t,\beta,\kappa)$ in this case as
\be\label{m33}
C(x,t,\beta,\kappa)=\frac{1}{\pi}\int^{x/2t}_{-\infty} (x-2t\lam) \ln|\varphi(\lam^2,\beta, \kappa)|\ d\lam+ \frac{1}{\pi} \int_{x/2t}^{\infty} (2t\lam-x)\ln|\varphi(\lam^2, \beta,\kappa)|\ d\lam\, . \ee
Using the expansion (\ref{m32}) one can see that the second integral in this equation is on the order of $O\left(e^{-((x/2t)^2-\beta)}\right)$, which for $x/2t$ outside of the immediate vicinity of $\sqrt{\beta}$, more precisely: $x/2t-\sqrt{\beta}>O(1/\sqrt{\beta})$, decreases exponentially in $\sqrt{\beta}$, since $(x/2t)^2-\beta >2 \sqrt{\beta} (x/2t- \sqrt{\beta})$. The same argument allows us to extend the upper limit of integration in the the first integral on the RHS of Eq.~(\ref{m33}) back to $+\infty$. Then,
the expansions (\ref{m311}) and (\ref{m32}) combined with the formulae
\[ e^{-\beta n}\int_0^{\sqrt{\beta}}e^{\lam^2 n}d\lam= \frac{1}{2n\sqrt{\beta}}+O\left(\frac{1}{\beta^{3/2}}\right),\ \ \
e^{\beta n}\int_{\sqrt{\beta}}^\infty e^{-\lam^2 n} d\lam= \frac{1}{2n\sqrt{\beta}}+O\left(\frac{1}{\beta^{3/2}}\right), \]
give the following estimate for this integral:
\be \frac{1}{\pi}\inti (x-2t\lam) \ln|\varphi(\lam^2,\beta, \kappa)|\ d\lam= -x\frac{2}{\pi\sqrt{\beta}} \sum_{n=1}^\infty \frac{ \cos(n\pi\kappa)+(-1)^{n+1}}{n^2}+O\left(\frac{1}{\beta^{3/2}}\right) .
\ee
Using the formulae (0.234) and (1.443) of \cite{GR}: $\sum_{k=1}^\infty(-1)^{n+1}/n^2=\pi^2/12$ and
$\sum_{k=1}^\infty\cos n\pi\kappa/n^2=\pi^2 B_2(\kappa/2)$, where  $B_2(x)=x^2-x+1/6$ is the second Bernoulli polynomial, and the fact that contribution of the region $\lambda >x/2t$ to the integral can be  neglected, we rewrite the previous result as
\be
\frac{1}{\pi}\inti|x-2t\lam|\ln|\varphi(\lam^2,\beta,\kappa)|\ d\lam
=-x\frac{\pi}{2\sqrt{\beta}}(1-\kappa)^2+O\left(\frac{1}{\beta^{3/2}} \right) ,\ \ \ (\beta\rightarrow\infty)\, .
\ee
Therefore, in the space-like region, we have
\be
C(x,t,\beta,\kappa)=-x\frac{\pi}{2\sqrt{\beta}}(1-\kappa)^2+O\left( \frac{1}{\beta^{3/2}}\right) ,\ \ \ (\beta\rightarrow\infty)\, .
\ee
{\bf Time-like region: ($x/2t<\sqrt{\beta}$)}. In this case, we begin by expressing $C(x,t,\beta,\kappa)$ as
\be
C(x,t,\beta,\kappa)=t\frac{4}{\pi}\int_0^{\infty}\lam \ln|\varphi(\lam^2, \beta,\kappa)|\ d\lam- \frac{2}{\pi}\int_0^{\frac{x}{2t}}(2\lam t-x) \ln|\varphi(\lam^2,\beta,\kappa)|\ d\lam \, .
\ee
As above, using the expansion (\ref{m311}), one can see that the second integral in this equation is on the order of $O(e^{-(\beta-(x/2t)^2)})$, which for $x/2t$ not too close to $\sqrt{\beta}$, more precisely: $\sqrt{\beta}-x/2t>O(1/\sqrt{\beta})$, decreases exponentially in $\sqrt{\beta}$, since $\beta-(x/2t)^2 >2 \sqrt{\beta} (\sqrt{\beta}- x/2t)$.   The, the expansions (\ref{m311}) and (\ref{m32}) and the calculations  similar to those in the space-like region give for the first integral:
\be
t\frac{4}{\pi}\int_0^{\infty}\lam\ln|\varphi(\lam^2,\beta,\kappa)|\ d\lam=-\pi t(1-\kappa)^2\, ,
\ee
The final result is
\be
C(x,t,\beta,\kappa)=-\pi t(1-\kappa)^2+O\left(e^{-(\beta-(x/2t)^2)}\right). \ee
%


\end{document}